\documentclass[pre,superscriptaddress,twocolumn,preprintnumbers,showpacs,amsmath,amssymb,10pt]{revtex4}
\usepackage{graphicx}
\usepackage{dcolumn}
\usepackage{subfigure}
\usepackage{hyperref}
\usepackage{ulem}
\usepackage{bm}
\usepackage{color}
\usepackage{latexsym}
\hypersetup{bookmarks,
            colorlinks,
           citecolor=red,
            filecolor=blue,
           urlcolor=blue,
           pdfpagemode=None}
\begin{document}
\title{Synchronization of oscillators with long range power law interactions}
\author{Debanjan Chowdhury}
\email{debanjanchowdhury@gmail.com}
\affiliation{Department of Physics, California Institute of Technology, Pasadena, CA 91125, USA}
\affiliation{Department of Physics, Indian Institute of Technology, Kanpur 208016, India}
\author{M.\ C.\ Cross}
\affiliation{Department of Physics, California Institute of Technology, Pasadena, CA 91125, USA}
\date{\today}%

\begin{abstract}
We present analytical calculations and numerical simulations for the synchronization of 
oscillators interacting via a long range power law interaction on a one 
dimensional lattice. We have identified the critical value of the power law exponent 
$\alpha_c$ across which a transition from a synchronized to an unsynchronized 
state takes place for a sufficiently strong but finite coupling strength in the large
system limit. We find $\alpha_c=3/2$. Frequency entrainment and phase ordering are
discussed as a function of $\alpha \geq 1$.
The calculations are performed using an expansion about the aligned phase state
(spin-wave approximation) and a  coarse graining approach.
We also generalize the spin-wave results to the {\it d}-dimensional problem. 
\end{abstract}
\maketitle
\section{Introduction}
Synchronization between a collection of oscillating objects is a common feature in a number of complex systems, 
such as the pacemaker cells in the heart, neurons in the nervous system, an array of Josephson junctions and rhythmic applause in a theater \cite{Pikovsky}.  At the same time, from a purely theoretical point of view, the phenomenon is interesting as perhaps the simplest examples of a collective response of driven dynamical systems, showing many features reminiscent of equilibrium phase transitions. Recent technological advancements in the fabrication of nanoelectromechanical systems (NEMS) promise large arrays of interacting nonlinear oscillators \cite{bargatin} that will provide good testing grounds for the theoretical predictions, as well as potential applications in sensitive detectors and precise frequency sources.

Winfree \cite{winfree} introduced a simple phase description of coupled oscillators, and he and Kuramoto \cite{kurabook} demonstrated synchronization for the case in which each oscillator is coupled equally to all the other oscillators (all-to-all coupling) using a mean field theory (for a review, see ref.\ \cite{acebron}): above a critical coupling strength depending on the distribution of the oscillator frequencies a finite fraction of the oscillators become entrained and oscillate at the same frequency, leading to a coherent signal in the summed response of the oscillators. On the other hand, for nearest neighbor coupling Daido \cite{daido} and Strogatz and Mirollo \cite{strogatz} showed that there is no macroscopic entrainment for a one dimensional chain. (By macroscopic we mean an $O(N)$ value for $N\to\infty$ oscillators.) In higher dimensions Strogatz and Mirollo further showed that any macroscopic entrained cluster must have the form of a sponge, i.e.\ any compact macroscopic entrained region contains ``holes'' of unentrained oscillators. The full behavior of the nearest neighbor model as a function of the dimension of the lattice is not completely understood, although based on approximate analytic arguments \cite{sakaguchi_1,daido} and numerics \cite{hong} it is conjectured that $d=2$ is the lower critical dimension for which macroscopic entrainment occurs.

We consider the phase model for a one-dimensional chain of oscillators with a strength of coupling between oscillators that falls off as a power law of their separation, and study how the synchronization behavior changes as a function of the power. This system was introduced by Rogers and Wille \cite{rogers}. They investigated the system numerically, in particular finding the the critical value of the power law for synchronization of a macroscopic system as a function of the coupling strength. They suggested a critical value $\alpha_{c}=2$ above which macroscopic synchronization would not occur for any finite coupling strength, no matter how large. Marodi et al.\ \cite{marodi} further investigated the system, focusing particularly on the question of complete entrainment in finite systems for small enough power laws. For an interaction power law $\alpha<1$, the sum of the interactions of a single oscillator with an infinite lattice of oscillators with aligned phases diverges for fixed coupling strength. In their study of this range of $\alpha$, Rogers and Wille \cite{rogers} chose a system size dependent normalization of the coupling strength to remove this divergence. This choice of normalization allows the investigation of the crossover to the all-to-all model of $N$ oscillators, where the coupling to each oscillator is scaled by $N^{-1}$ so that the synchronization transition occurs at finite coupling constant. On the other hand, Marodi et al.\ \cite{marodi} argued that in physical systems of interest, the interaction strength would not be expected to scale with the system size, and they investigated the model without scaling the coupling constant with system size A major focus of their study was how the size of the system needed for complete entrainment then depends on the interaction power law and the coupling strength. 

A power law interaction is of interest both experimentally and theoretically. This type of interaction should be relevant to some implementations of nanomechanical arrays, since both electrostatic interactions between charges or dipoles on the devices and elastic interactions through the supporting substrate may lead to such long range interactions. Interactions falling off as a power law have also been used to model the complex long range connectivity of neurons  \cite{pnas}. Radicchi and Meyer-Ortmanns \cite{radicchi} have recently investigated the case of a single pacemaker oscillator with a different frequency coupled through a power law interaction to many identical oscillators. From a purely theoretical point of view, the long range coupling with an appropriate normalization factor also allows one to interpolate between the extreme cases of all-to-all and nearest neighbor coupling. Finally, the model with power law interactions allows us to assess  the accuracy of various analytic approximation schemes by comparison with large numerical simulations.

In this paper we present a more systematic investigation of the power law model, using analytic perturbation techniques analogous to spin-wave theory in magnets and cluster arguments following ref.\ \cite{strogatz}, as well as numerical simulations on larger systems than in previous work. We study the range of interaction power laws $\alpha > 1$, and the question of whether macroscopic synchronization may exist for a large number of oscillators, and the nature of the synchronized state, as a function of $\alpha$. We emphasize that for this range of interaction power laws, the different choice of normalization of the coupling strength used in refs.\  \cite{rogers,marodi} results only in a finite multiplicative factor of the coupling strength, and does not change any of the qualitative results. Only for $\alpha\le 1$ is the choice of normalization critical to the questions being addressed.

The outline of the paper is as follows. In Section II we introduce the model and the diagnostics we use to quantify its behavior. We then describe three approaches to understand the behavior. In Section III we perform an expansion about the aligned-phase state analogous to the spin-wave approximation in magnetic systems. In Section IV we coarse grain the system by summing over blocks of oscillators, following the method used by Strogatz and Mirollo \cite{strogatz} in their discussion of the nearest neighbor model. Numerical simulations on large systems of up to 16384 oscillators are described in Section V. In Section VI we bring together the results, compare with previous work on the long range model \cite{rogers,marodi}, and conclude.

\section{Model and Diagnostics}

In the simplest model of a population of mutually interacting oscillators with different frequencies, each oscillator is reduced to a single phase degree of freedom which evolves at a rate determined by its intrinsic frequency and its interactions with the other oscillators proportional to the sine of the phase differences \cite{kurabook}
\begin{equation}
\dot\theta_{j}=\omega_{j}+\sum_{i\neq j}K_{ij}\sin(\theta_{i}-\theta_{j}).
\label{kuragen}
\end{equation}
The all-to-all model
($K_{ij}=K/N$) and the short range model, where only the nearest neighbors 
interact with each other ($K_{ij}=K$ for nearest neighbors, zero otherwise) have been studied in great detail \cite{acebron}.
 
In this paper, we consider oscillators with a power law coupling that varies as $K_{ij}=K/r_{ij}^{\alpha}$, where 
$r_{ij}$ is the distance between the oscillators at site $i$ and $j$.
The model is defined by the equations of motion
\begin{equation}
\dot\theta_j=\omega_j + K\sum_{s=1}^{N-1}\frac{1}{s^{\alpha}}\left[\sin(\theta_{j+s}-\theta_{j})+\sin(\theta_{j-s}-\theta_{j})\right].
\label{longmod}
\end{equation}
Here, $\theta_j\ (1\leq j\leq N)$ is the phase of the $j^{th}$ oscillator and $\omega_j$ are the corresponding intrinsic frequencies, assumed to be independent random variables with distribution $g(\omega)$. Without loss of generality, $g(\omega)$ can be chosen such that $\langle \omega_{j}\rangle =0$ and, for bounded distributions, $\langle \omega_j^2\rangle =1$. The parameter $K$ sets the strength of the coupling, with $\alpha$ the exponent for the power-law decay of the interactions.
We use periodic boundary conditions so that $j+N\equiv j$. 

We are interested in the synchronization of the oscillators to one another. A number of different criteria for synchronization can be introduced.

We will use the term entrainment to denote oscillators that are evolving with the same frequency. More precisely, we will define oscillators $i,j$ as {\it entrained} if there are no $2\pi$ phase slips over arbitrarily long time evolution after initial transients have died out
\begin{equation}
\text{Entrainment:}\quad|\Delta\theta_{ij}(t_{0}+T)-\Delta\theta_{ij}(t_{0})|<2\pi
\end{equation}
with $\Delta\theta_{ij}=\theta_{i}-\theta_{j}$. Note that this is stricter than simply requiring the long-time mean frequencies $\bar{\omega}_{i}=(\theta_{i}(t_{0}+T)-\theta_{i}(t_{0}))/T$ to be equal, since for example phase diffusion $|\Delta\theta_{ij}(t)|\sim t^{1/2}$ would be consistent with the latter condition but not the former.
A measure of the presence of entrainment over the whole system or frequency order is the Edwards-Anderson order parameter
\begin{equation}
\Psi_{EA}=\lim_{t-t_{0}\to\infty}\frac{1}{N}\left|\sum_{j=1}^{N}e^{i(\theta_{j}(t)-\theta_{j}(t_{0}))}\right|.
\label{define EA}
\end{equation}
For a {\it fully entrained} state all of the oscillators evolve with the same frequency, which will be the mean of the frequency distribution, and $\Psi_{EA}=1$ . This is a particularly simple state: a periodic solution (limit cycle) in general and a time independent solution (fixed point) after setting the mean frequency to zero.

Another measure of synchronization is {\it phase order} giving the average alignment of the phases of the oscillators. The degree of phase order over the system is quantified by the magnitude of the phase 
order parameter $|\Psi_{ph}(t)|$ at any time with
\begin{equation}
\Psi_{ph}(t)=\frac{1}{N}\sum_{j=1}^{N}e^{i\theta_{j}(t)}.
\label{define OP}
\end{equation}
The signal measured in an experiment that sums the signals from the individual oscillators is proportional to $\operatorname{Re}\Psi_{ph}(t)$.
A state with perfectly aligned phases has $|\Psi_{ph}|=1$.
In principle, any time dependence of $\Psi_{ph}$ is possible, but for the phase model we expect the synchronized motion for large $N$ to be close to periodic. A fully entrained state (all oscillators entrained), for example, will give a periodic $\Psi_{ph}$, but typically with $|\Psi_{ph}|<1$ since the phases will not be fully aligned. We will usually investigate the time average magnitude $\langle |\Psi_{ph}(t)|\rangle_{t}$ after some time $t_{0}$ to allow transients to decay. We also introduce the {\it phase correlation function} given by
\begin{equation}
C_{ij}=\langle e^{i(\theta_{i}-\theta_{j})}\rangle,
\label{phase correlation function}
\end{equation}
with the average extending over time and over the lattice $i,j$ for fixed separation $i-j$.

\section{Spin-wave Analysis}
We first  carry out a spin-wave (SW) type analysis of this model. Such an approach has been applied
to the short range model earlier \cite{kuraprog}. This approach studies the small deviations from a state of aligned phases, and investigates the consistency of this assumption.

\subsection{Preliminaries}
\label{section preliminaries}
For large enough coupling, we might anticipate a fully entrained state where each oscillator evolves with the same frequency, and where the phase differences $\Delta\theta_{ij} =\theta_{i}-\theta_{j} $ between the interacting oscillators are small. The common frequency will be the mean of the frequency distribution, which we have set to zero,
and so the fully entrained state is time independent. If the phase differences $\Delta\theta_{ij}$ are small, the sine functions appearing in the interaction term may be linearized. Introducing the Fourier modes
\begin{equation}
\theta_{m}=\sum_{q}{\tilde{\theta}}_{q}e^{iqm},
\end{equation}
with $q=2n\pi/N$ with $n$ integral and the sum running over the first Brillouin zone
$-\pi < q \leqslant\pi$, yields
\begin{equation}
\dot{\tilde{\theta}}_{q}={\tilde{\omega}}_{q}-K(q){\tilde{\theta}}_{q},
\label{theta q}
\end{equation}
with the interaction kernel
\begin{equation}
K(q)=2K\sum_{s=1}^{N-1}\frac{1-\cos qs}{s^{\alpha}}.
\label{Kq}
\end{equation}
Since $K(q)$ is positive, each mode relaxes exponentially to a steady state determined by the Fourier transform of the random frequencies $\tilde{\omega}_{q}$: we therefore investigate the properties of this steady state, which corresponds to the fully entrained state. Solving for the steady state of Eq.\ (\ref{theta q}) and using $\langle \omega_j^2\rangle =1$,
we obtain for the mean square phase difference
\begin{equation}
\langle|\Delta\theta_{ij}|^{2}\rangle=\frac{1}{N}\sum_{q\ne0}|K(q)|^{-2}(1-\cos[q(i-j)]).
\label{fluc}
\end{equation}

\begin{figure}
\begin{center}
\includegraphics[width=1.0\columnwidth]{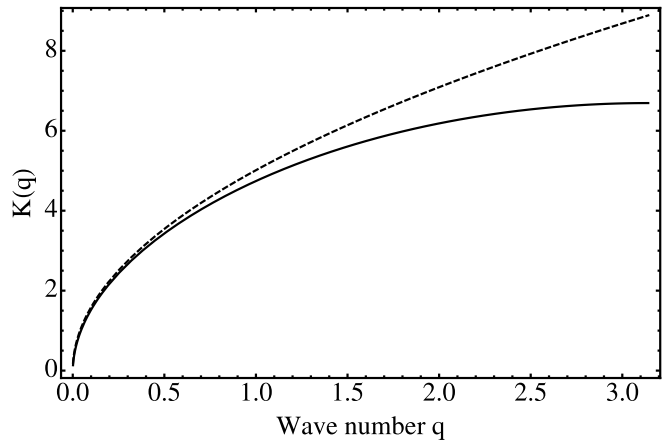}
\end{center}
\caption{The interaction kernel $K(q)$ (solid) and the continuum approximation
Eq.\ (\ref{intcont1D}) (dashed) for $\alpha=3/2$ and $N=4096$. The inset shows the same curves on a log-log scale.}
\label{kernel}

\end{figure}

Important issues are the behavior of $\langle|\Delta\theta_{ij}|^{2}\rangle$
for large separations $i-j$, which depends on the small $q$ terms in the sum in Eq.\ (\ref{fluc}), and the possible divergence of $\langle|\Delta\theta_{ij}|^{2}\rangle$
for any $i-j$ due to the vanishing of $K(q)$ for small $q$. For small $q$ and large $N$, with $q\gg N^{-1}$,
we can evaluate $K(q)$ in the  
continuum limit by replacing the sums by integrals with the appropriate density
of states
\begin{equation}
K(q)\simeq2K\int_{0}^{\infty}ds\frac{1-\cos{qs}}{s^{\alpha}}.
\label{intcont1D}
\end{equation}
The integral can be evaluated for $1<\alpha<3$ to give
\begin{equation}
K(q)=Kcq^{\alpha -1},\label{kqint}
\end{equation}
where
\begin{equation}
c=2\sin(\pi\alpha/2)\alpha\Gamma(-\alpha),\label{c}
\end{equation}
with $\Gamma$ the Euler Gamma function, except at precisely $\alpha=2$ where $K(q)=\pi K q$.
(This is also the limit of the general expression for $\alpha\to 2$.)
In Fig.\ {\ref{kernel}}, we show a comparison between the exact $K(q)$ evaluated from Eq.\ (\ref{Kq}) and the continuum approximation Eq.\ (\ref{kqint}) for $\alpha=3/2$ and $N=4096$, demonstrating the accuracy of the approximation for $N^{-1}\ll q\ll 1$, but over a range that is restricted by the finite size effects. The integral Eq.\ (\ref{intcont1D}) and the sum Eq.\ (\ref{Kq}) for
$N\rightarrow\infty$ both diverge for $\alpha<1$.

For $N\rightarrow\infty$, the mean 
square phase difference Eq.\ (\ref{fluc}) can be evaluated replacing the sum by an integral 
\begin{equation}
\langle|\Delta\theta_{ij}|^{2}\rangle=\frac{1}{\pi}\int_{0}^{\pi}dq\,|K(q)|^{-2}(1-\cos[q(i-j)]).
\label{Delta continuum}
\end{equation}
For small $q$ the kernel behaves as $K(q)\sim q^{\alpha-1}$ and the integral Eq.\ (\ref{Delta continuum}) diverges from the small $q$ behavior for $\alpha>5/2$. We interpret the divergence of $\langle|\Delta\theta_{ij}|^{2}\rangle$ to signal the onset of phase slips, and the breakdown of the assumption of a time independent, fully entrained, solution. This argument therefore suggests that there is no fully entrained state as $N\to\infty$ for $\alpha>5/2$.

We can also evaluate the large distance behavior of $\langle|\Delta\theta_{ij}|^{2}\rangle$. For $i-j\rightarrow\infty$ the function $\cos[q(i-j)]$ is rapidly oscillating as a function of $q$ and this term in Eq.\ (\ref{Delta continuum}) averages to zero. The remaining integral diverges, again because of the small $q$ behavior, for $\alpha>3/2$, but is finite for $\alpha<3/2$.
This implies that for $\alpha<3/2$ and large enough coupling strength $K$, the phase difference $\theta_{i}-\theta_{j}$ is small even for large $i-j$, suggesting a state with long range phase order, as well as entrainment. On the other hand for $3/2<\alpha<5/2$ the argument suggests an entrained state without long range phase order.

In the following sections we study the properties of the states in more detail, evaluating the phase correlation function and the order parameter in the spin-wave approximation as a function of $\alpha$.

\subsection{Correlation Function}
\label{section correlation function}
The phase correlation function $C_{ij}=\langle\cos(\theta_{i}-\theta_{j})\rangle$ is given by
\begin{equation}
C_{ij}=\langle e^{i(\theta_{i}-\theta_{j})}\rangle=e^{-\langle|\Delta\theta_{ij}^{2}|\rangle},
\label{correlation function}
\end{equation}
with $\langle|\Delta\theta_{ij}^{2}|\rangle$ obtained from Eq.\ (\ref{Delta continuum}). We now evaluate this expression for different ranges of $\alpha$. The results are summarized in Eq.\ (\ref{C cases}) below.

For $\alpha<3/2$ we write Eq.\ (\ref{Delta continuum}) as
\begin{equation}
\langle|\Delta\theta_{ij}|^{2}\rangle=\langle|\Delta\theta_{\infty}|^{2}\rangle-\frac{1}{\pi}\int_{0}^{\pi}dq\,|K(q)|^{-2}\cos[q(i-j)],
\label{Delta split 3/2}
\end{equation}
with
\begin{equation}
\langle|\Delta\theta_{\infty}|^{2}\rangle=\frac{1}{\pi}\int_{0}^{\pi}dq\,|K(q)|^{-2},
\label{Delta infty}
\end{equation}
the finite $i-j\rightarrow\infty$ asymptotic value. In the remaining integral in Eq.\ (\ref{Delta split 3/2}), the contribution from the range $q>(i-j)^{-1}$ is small since the cosine term is rapidly oscillating here, so that for large $i-j$ the small $q$ expression (\ref{kqint}) can be used for $K(q)$ and the upper limit replaced by $\infty$. This gives for large $r=i-j$
\begin{eqnarray}
C_{r}&=& C_{\infty} \exp\left[-\frac{1}{K^2 \xi(\alpha)r^{3-2\alpha}}\right],\\
&\simeq &  C_{\infty}\left[1-\frac{1}{K^{2}\xi(\alpha)r^{3-2\alpha}}\right],
\label{C < 3/2}
\end{eqnarray}
where
\begin{equation}
\xi(\alpha ) = \pi c^{2}/\sin(\pi\alpha)\Gamma(3-2\alpha),\\
\end{equation}
with the constant $c$ defined in Eq.\ (\ref{c}), and
\begin{equation}
C_{\infty}=e^{-\langle|\Delta\theta_{\infty}|^{2}\rangle}.
\label{C infty}
\end{equation}
Note that $\xi(\alpha)$ is negative for $\alpha<3/2$, so that Eq.\ (\ref{C < 3/2}) represents
correlations growing as a power law above the large distance value as $r$ is decreased.

For $\alpha>3/2$, the full integral in Eq.\ (\ref{Delta continuum}) is dominated by the small $q$ region, so that again the expression Eq.\ (\ref{kqint}) can be used for $K(q)$ and the upper limit replaced by $\infty$. This gives
\begin{equation}
\langle|\Delta\theta_{ij}|^{2}\rangle = \frac{|i-j|^{2\alpha-3}}{K^{2}\xi(\alpha)},
\end{equation}
and the result for the correlation function
\begin{equation}
C_{r}=e^{-r^{2\alpha-3}/K^2\xi(\alpha)}.
\end{equation}
Two special cases of note are $\alpha = 3/2$ where there are power law correlations
\begin{equation}
C_{r}\propto r^{-1/8\pi^2K^{2}}
\end{equation}
(we have not evaluated the proportionality constant),
and $\alpha=2$ where the correlations are simply exponential
\begin{equation}
C_{r}=e^{-r/2\pi^2 K^2}.
\end{equation}

In summary, we find the following result for the correlation function $C(r)$ at large separations $r$
and in the limit $N\to\infty$
\begin{eqnarray}
C(r)
\begin{cases}
=1,~~~~~~~~~~~~~~~~~~~~~~~~~~~~~~~~~~~~~~\alpha\leq1,\\
= C_{\infty} \exp{[-1/K^2 \xi(\alpha)r^{3-2\alpha}]}, ~~1<\alpha<3/2,\\
\propto r^{-1/8\pi^2K^{2}},~~~~~~~~~~~~~~~~~~~~~~~~~~\alpha=3/2,\\
=\exp{[-r^{2\alpha-3}/K^2\xi(\alpha)]}  , ~~~3/2<\alpha(\neq2)<5/2, \\
=\exp{[-r/2\pi^2 K^2]}~~~~~~~~~~~~~~~~~~~~\alpha=2.\
\end{cases}
\label{C cases}
\end{eqnarray}
Thus as $\alpha$ is varied, the spin-wave approximation predicts long range phase order for $\alpha<3/2$,
and then as $\alpha$ increases the correlation function crosses over from a power law decay
to stretched exponential, exponential, and then super-exponential. Similar results have been obtained using the spin-wave approximation for the classical XY model \cite{tasaki} 
as the dimension is varied continuously between $1<d<2$. In that work they were able to show by other methods that the stretched exponential prediction was an artifact of the spin-wave approximation, and that correlations were bounded by a simple exponential fall off. We do not know if a similar result might apply to the present model, although we find some confirmation of the stretched exponential behavior in the numerical simulations presented in Section \ref{num} below.


\subsection{Order Parameter}
\label{Section Order Parameter}

\begin{figure}[htf]
\begin{center}
\includegraphics[width=3.0in]{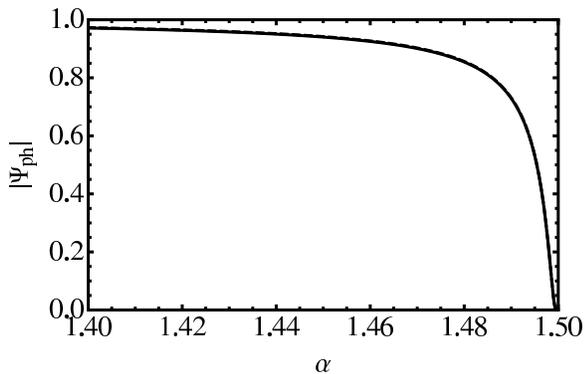}
\caption{Phase order parameter $|\Psi_{ph}|$ for an infinite one dimensional chain calculated using the spin-wave approximation as a function of the power-law for the decay of interactions $\alpha$: solid - full calculation
from Eqs.\ (\ref{Kq}) and (\ref{Delta continuum}); dashed - approximation given by Eq.\ (\ref{Psi approx}). The coupling strength is $K=1.0$}
\label{Psi spinwave}. 
\end{center}
\end{figure}

An important measure of the coherence of a set of oscillators
is the phase order parameter Eq.\ (\ref{define OP}). For an infinite system, the order parameter may be obtained from the asymptotic value of the correlation function
\begin{equation}
\label{OP infinite}
|\Psi_{ph}|=\sqrt{C_{\infty}}.
\end{equation}
with $C_{\infty}$ given by Eq.\ (\ref{C infty}) and then Eq.\ (\ref{Delta infty}). The full expression must be evaluated numerically. However we can get an analytic approximation using the approximate expression (\ref{kqint}) for $K(q)$. This should be a good approximation for $\alpha$ near $3/2$ where the small $q$ range dominates the integral. This gives the estimate
\begin{equation}
|\Psi_{ph}|\sim\text{Exp}\left(-\frac{1}{2 \pi  K^2 [2 \text{Sin}(\pi  \alpha /2) \alpha  \Gamma(-\alpha)]^2} \frac{\pi ^{3-2 \alpha }}{3-2 \alpha }\right)\ .
\label{Psi approx}
\end{equation}
This approximation is compared with the result calculated numerically using the full expression for $K(q)$
given by the sum Eq.\ (\ref{Kq}) for $N\to\infty$ in Fig.\ \ref{Psi spinwave}.
For $\alpha\to 3/2$ Eq.\ (\ref{Psi approx}) becomes
\begin{equation}
|\Psi_{ph}|\to\text{Exp}\left[-\frac{1}{16 \pi^{2}  K^2 (3-2 \alpha ) } \right]\ .
\end{equation}
Thus, no matter how large $K$ is, for $\alpha$ close enough to 3/2 the order parameter decreases,
and goes to zero at $\alpha=3/2$ in the spin wave approximation.

The magnitude of the order parameter in a finite system can be calculated from the 
correlation function
\begin{equation}
\label{OP}
|\Psi_{ph}|=\sqrt{\frac{1}{N^{2}}\sum_{i,j=1}^{N}C_{ij}}.
\end{equation}
We have evaluated this expression, performing the discrete sums for the spin wave approximation. These results will be compared with simulations on the dynamical equations in Section \ref{num} below.

It is also of interest to get a more approximate estimate of the order parameter in a finite system for $\alpha$ near 3/2. If the correlations remain sizable over the whole system we may estimate the order parameter as $|\Psi_{ph}(N)|\sim\sqrt{C(N/2)}$. (This estimate fails for large enough $N$, or when the order parameter gets small, so that the correlations become small over much of the system.) Then we approximate $C(N/2)$ from Eq.\ (\ref{Delta continuum}): we use the power law form Eq.\ (\ref{kqint}) for $K(q)$ which is good for the small $q$ region that dominates the integral for $\alpha\simeq 3/2$; over most of the range integration $\beta\pi/N < q <\pi$ (with $\beta$ an $O(1)$ number) we argue that the cosine term oscillates rapidly to zero; and over the remainder of the range $0 < q < \beta\pi/N$ we have $1-\cos(qN/2)\propto q^{2}$ so that the integrand becomes small. This gives the estimate
\begin{equation}
\langle|\Delta\theta_{N/2}|^{2}\rangle\sim \frac{\pi^{3-2\alpha}}{\pi c^{2} K^{2}(3-2\alpha)}\left[1-\left(\frac{\beta}{N}\right)^{3-2\alpha}\right],\label{Delta theta N/2}
\end{equation}
and then
\begin{equation}
|\Psi_{ph}(N)|\sim e^{-\langle|\Delta\theta_{N/2}|^{2}\rangle /2}.
\label{Psi N estimate}
\end{equation}
Note that the expression behaves smoothly through $\alpha=3/2$, and at this value reduces to
\begin{equation}
|\Psi_{ph}(N)|_{\alpha=3/2}\sim(N/\beta)^{-\tfrac{1}{16\pi^{2} K^{2}}}.
\end{equation}
For $K=1$ these expression show a very slow scaling to the $N\to\infty$ limits for $\alpha$ near 3/2: for example to achieve $|\Psi_{ph}(N)|_{\alpha=3/2}<0.5$ requires $N>3\times10^{47}$ (setting $\beta=1$).

\subsection{Self Consistency}

We might worry, if the order parameter is becoming small as $\alpha\to 3/2$, that the spin-wave approximation breaks down in this vicinity due to the failure of the linearization of $\sin(\theta_{i}-\theta_{j})$.

To estimate the typical size of $|\theta_{i}-\theta_{j}|$ for the interacting oscillators, we calculate the average of the correlation function {\it weighted by the strength of the interaction} at that separation
\begin{equation}
\bar{C}=\frac{\int_{1}^{\infty}C_{r}r^{-\alpha}dr}{\int_{1}^{\infty}r^{-\alpha}dr}.
\end{equation}
Evaluating the integral for the power law correlations at $\alpha=3/2$ given by Eq.\ (\ref{C cases}), and approximating for large $K$ gives
\begin{equation}
\bar{C}\simeq1-\frac{1}{4\pi^{2} K^{2}}.
\end{equation}
Thus for large enough $K$, the average phase correlation important in the interactions is close to unity,
as assumed in the spin wave approximation, even though the order parameter is zero. For example,  $\bar{C}>0.75$ requires $K>\pi^{-1}$.

For smaller values of $\alpha$, or in a finite system, the correlations are enhanced, and so the approximation should be even better in these cases.

\subsection{General Dimension}

The results of the spin-wave calculation can be generalized to $d$-dimensional lattices.
The corresponding results for the correlation function are: perfect phase ordering with $C({\bf r})=1$ for $\alpha\leq d$; entrainment
with long range phase order for $\alpha<3d/2$;  
a power-law fall off of the phase correlation function at $\alpha=3d/2$; and a cross over to 
exponential decay of correlations at $\alpha=(3d+1)/2$ via stretched exponentials. 

\section{Block Sums}

\subsection{Preliminaries}

In this section, we will analyze the long range problem by coarse 
graining the one dimensional chain into block 
oscillators. It is useful to coarse grain the chain into blocks by summing the
equations of motion Eq.\ (\ref{longmod}) for all $\theta_{i}$ in the block, since in this 
way, the internal interactions within a given block cancel with each other. This means one can look at the interaction of the block with the rest of the 
chain. In the short range model, this turns out to be especially useful, since
the interaction of a block with the rest of the chain includes the
surface terms only. In the long range model the situation is more 
complex, because the oscillators within a block interact with all the oscillators in the 
rest of the chain.
In this section, for simplicity we use open boundary conditions rather than periodic ones.

Before we discuss the results for the long range model, let 
us recapitulate the results obtained by Strogatz and Mirollo \cite{strogatz} 
for the nearest neighbor model.
A key result which we will try and generalize 
to the long range model is the following. They find that
it is impossible to have a macroscopic synchronized cluster in one dimension for finite
values of the coupling constant. In higher
dimensions, any macroscopic cluster takes the form of a sponge, i.e.\ the 
cluster is riddled with holes, which correspond to unsynchronized oscillators.

We write the basic equations (\ref{kuragen}) in the form
\begin{equation}
\dot\theta_{j}-\omega_{j}=\sum_{i\neq j}K_{ij}\sin(\theta_{i}-\theta_{j}).
\label{kuragenrw}
\end{equation}
The average frequency $\tilde\omega_j$  of the $j$th oscillator
is defined as $\tilde\omega_j=\lim_{t\to\infty}(\theta_j(t)-\theta_{j}(0))/t$.
We define a block ${\cal{S}}$ as a contiguous segment of the chain of oscillators. For a synchronized block the oscillators must all have the same average frequency, $\tilde\omega_j=\tilde\omega$ for all $j\in{\cal{S}}$. We now sum the equations over a synchronized block ${\cal{S}}_{k}$ of $M$ oscillators entrained at frequency $\tilde\omega_{k}$.

In analyzing the equation in terms of block sums we need the properties of two quantities: the frequency sums also known as the accumulated randomness; and the interaction sums.

Summing the time averaged equations of motion (\ref{kuragenrw}) over the block ${\cal{S}}_{k}$ will give on the left hand side the quantity
$M\tilde\omega_{k}-Y_{k}(M)$ with
\begin{equation}
Y_{k}(M)=\sum_{j\in{\cal{S}}_k}\omega_{j},
\label{Y km}
\end{equation}
the accumulated randomness.
We will also use
\begin{equation}
y_{k}(M)=M^{-1}Y_{k}(M)=M^{-1}\sum_{j\in{\cal{S}}_k}\omega_{j}.
\label{y_k}
\end{equation}
A key point in the arguments below is that the {\it support} of possible values of $y_{k}$ is the same as the support of the individual  frequencies $\omega_{j}$, but for large $M$ the {\it typical} value of $y_{k}$ will scale as $M^{-1/2}$ for frequency distributions with a finite variance. For example, for a bounded frequency distribution, there is some probability (very small for large $M$) that each $\omega_{j}$  in the block will have the maximum possible value $\omega_{\text{max}}$, and then $y_{k}$ will take on the value $\omega_{\text{max}}$. On the other hand, since for large $M$ the central limit theorem means that $y_{k}$ is given by a Gaussian distribution with standard deviation scaling as $M^{-1/2}$, the typical values of $y_{k}$ (those with nonzero probability for $M\to\infty$) are of order $M^{-1/2}$. Obviously, similar remarks apply to $Y_{k}$ after including an additional factor of $M$.

The second quantity of interest is the summed interaction of the block ${\cal{S}}_{k}$ with other oscillators in the chain. Consider first the summed interaction of this block, with a second block ${\cal{S}}_{p}$ of size $\bar{M}$.
The largest possible interaction sum, when all the phases within each block are aligned is
\begin{equation}
I_{kp}=\sum_{i\in{\cal{S}}_k,j\in{\cal{S}}_p}K_{ij}.
\label{I_kp max}
\end{equation}
Using the bound
\begin{equation}
\sum_{n=p}^{q}f(n)<\int_{p-1/2}^{q+1/2}f(x)\,dx
\end{equation}
for any function $f(x)$ with positive curvature, we can bound the interaction sum for $1<\alpha<2$
for the two blocks separated by $D$ oscillators\begin{equation}
\begin{split}
I_{kp} < 
\bar{K}[(M+D)^{2-\alpha}+(\bar{M}+D)^{2-\alpha}
\\ -  (M+\bar{M}+D)^{2-\alpha}-D^{2-\alpha}],
\end{split}
\label{I_kp bound upper}
\end{equation}
with
\begin{equation}
\bar{K}=\frac{K}{(2-\alpha)(\alpha-1)}.
\end{equation}
Similarly, using the bound
\begin{equation}
\sum_{n=p}^{q}f(n)>\int_{p}^{q}f(x)\,dx
\end{equation}
for any monotonically decreasing function gives for $1<\alpha<2$
\begin{equation}
\begin{split}
I_{kp} > 
\bar{K}[(M+D)^{2-\alpha}+(\bar{M}+D)^{2-\alpha}
\\ -  (M+\bar{M}+D-1)^{2-\alpha}-(D+1)^{2-\alpha}].
\end{split}
\label{I_kp bound lower}
\end{equation}

We will use various limits of these expressions. For example, for a block of size $M$ interior to an infinite chain and interacting with the remainder of the chain we use $D=0,\bar{M}\to\infty$ for the interaction with the infinite chain in either direction, to find for $1<\alpha<2$
\begin{equation}
I_{kp} \sim 2\bar{K}M^{2-\alpha}
\label{M bound}
\end{equation}
(both bounds lead to the same expression in this limit).
The essence of the block-sum arguments is to compare the interaction sum, scaling as $M^{2-\alpha}$, with either the range of possible values of the frequency sum, scaling as $M$, or the typical value, scaling as $M^{1/2}$. This will yield important changes of behavior at $\alpha=1$ and $\alpha=3/2$. For $\alpha>2$ the interaction sums between large blocks are independent of the sizes of the blocks, and the results essentially  reduce to those of the model with nearest neighbor interactions.

\subsection{Impossibility of Macroscopic Clusters for Finite Coupling Strength and $\alpha > 1$}
\label{smclust}
In this section, we will obtain the regime of $\alpha$ where it is impossible to have a macroscopic sized synchronized block. More precisely, define the probability $P(N,K,f)$ that there exists one or more contiguous blocks ${\cal{S}}$ containing at least $fN$ oscillators with $f$ some finite fraction. Then $P(N,K,f)$ is zero for $N\to\infty$ and $K$ finite. The approach closely follows the one Strogatz and Mirollo \cite{strogatz} used for the short range model.

Let us suppose that such a block ${\cal{S}}$ is made up of 
synchronized oscillators at a frequency $\tilde\omega$. We now divide the 
block into $R$ nonoverlapping segments, ${\cal{S}}_k$ of length $m$ each. Thus $R=fN/m$ and $k$ varies from $1$ to $R$. Note that $m$ is an integer, sufficiently large but finite
as $N\rightarrow\infty$ and $R$ is ${\cal{O}}(N)$. Summing the time-averaged equations of motion (\ref{kuragenrw}) over the sub-block ${\cal{S}}_k$ and bounding the interaction sum as described above, for the sub-block ${\cal{S}}_k$ to be part of the synchronized block, we must have, for  $N>>m$ and $1<\alpha<2$,
\begin{equation}
|\tilde\omega-y_k| < KF_{\alpha}(m),
\label{clust}
\end{equation}
with, using Eq. (\ref{I_kp bound upper}) for $M=m,D=0,\bar M\to\infty$,
\begin{equation}
F_{\alpha}(m)=\frac{2}{(\alpha-1)(2-\alpha)}\frac{1}{m^{\alpha-1}} .
\end{equation}
For $\alpha>2$ the interaction sums converge for large $m,N$ so that $F_{\alpha}(m)=f/m$ with
$f$ an O(1) constant. In the latter case the expression reduces to the one for the short
range model. For $\alpha<1$, $F_{\alpha}(m)$ diverges as $N\rightarrow\infty$.

Now we argue that for $\alpha>1$ we can choose $m$ sufficiently large but finite such that the probability $p$ of Eq.\ (\ref {clust}) being satisfied, for block $k$ and  a given $\tilde\omega$, is less than unity. This follows, because there is some nonzero probability of finding any value of $y_{k}$ over the support of the $\omega_{j}$ probability distribution, whereas the right hand side may be made as small as we choose by choosing $m$ sufficiently large. Lemma 3.1 given in \cite{strogatz} makes this argument precise. It then follows that the probability that the result is satisfied for {\it all} $R$ sub-blocks is $p^{R}$, and scales to zero as $N\to\infty$ as $O(e^{-cN})$ with $c$ some positive constant (remember $R$ is $O(N))$. Since the block $\cal{S}$ can be located at a number of different locations which is certainly less than $N$, this means that the probability for macroscopic blocks satisfies $P(N,K,f) < O(Ne^{-cN})$ which tends to zero as ${N\to\infty}$.
 
Thus it is impossible to have a contiguous macroscopic block,
containing an $O(1)$ fraction of the oscillators locked to a common frequency, when $\alpha>1$. In any macroscopic segment of the chain there will always be (i.e.\ probability one as $N\rightarrow\infty$) some finite blocks of ``runaway'' oscillators that are desynchronized from their neighbors. 
This is the same result as in the nearest neighbor model: for the question of the formation of
finite blocks of unsynchronized oscillators, the power law interactions do not change the conclusions unless the interaction with a single oscillator can be infinite (as is the case for $\alpha<1$). 

\subsection{Synchronization of Large Separated Blocks for $\alpha<3/2$}
\label{mutual}
For the nearest neighbor model, the result analogous to the one of the previous section is sufficient to show that for a one dimensional chain there will not be a macroscopic number of synchronized oscillators for finite coupling, since the unsynchronized blocks effectively cut the chain into noninteracting pieces. However, for the long range model it is perhaps possible, even for a finite $K$ and $\alpha>1$, to have a partially entrained state with a macroscopic number of oscillators having the same frequency. This would correspond to the system breaking up into disconnected blocks which however synchronize via the long range interaction across the unsynchronized oscillators. In this section we analyze the mutual interaction between two distant blocks, which are separated by blocks of unsynchronized oscillators, and investigate their possible synchronization. It is more difficult to prove the existence of synchronization, rather than its absence, and the argument we present is less rigorous than in the previous section.
  
Let us consider the two large blocks, ${\cal{S}}_k$ and ${\cal{S}}_p$, of size $M$ each, where $M>>1$, and $\alpha$ in the range $1<\alpha<2$. From the general expression Eq.\ (\ref{I_kp bound lower}) we obtain the following results. For a separation $D\ll M$ (e.g. $D$ finite and $M=O(N)$ for $N\to\infty$)\begin{equation}
I_{kp}>2\bar{K}(1-2^{1-\alpha})M^{2-\alpha}
\end{equation}
{\it independent} of the separation $D$ with $O(D/M)$ corrections. Also, for blocks $k,p$ separated by $p-k-1$ blocks of size $M$ 
\begin{equation}
I_{kp}>c_{pk}\bar{K}M^{2-\alpha}
\end{equation}
with $c_{pk}$ an $O(1)$ number
\begin{equation}
c_{pk}=2(p-k)^{2-\alpha}-(p-k+1)^{2-\alpha}-(p-k-1)^{2-\alpha}.
\end{equation}
In both cases, the lower bound on the maximum interaction sum scales as $M^{2-\alpha}$.
(It can be shown the upper bound scales in the same way).

On the other hand the frequency sums $Y_{k},Y_{m}$ Eq.\ (\ref{Y km}) are described, for large $M$, by independent Gaussian distributions with standard deviation scaling as $M^{1/2}$. The typical difference between the frequency sums will also scale as $M^{1/2}$. This means that two large, fully, aligned blocks, each of $M$ oscillators, separated by finite blocks of unsynchronized oscillators, or even by $O(M)$ oscillators, will typically synchronize for $\alpha<3/2$ for coupling strengths $K>O(M^{\alpha-3/2})$. We would expect this result to extend to blocks that are not fully aligned, provided each block has  a nonzero value of the phase order parameter $\Psi_{k},\Psi_{p}$, since we would expect the interaction sum to be reduced by a factor of about $|\Psi_{k}\Psi_{p}|$. Thus, for $\alpha < 3/2$ the small blocks of unsynchronized oscillators, necessarily present by the arguments of the previous section, do not necessarily act to break the chain into finite lengths of synchronized oscillators, and macroscopic synchronization is possible.

For $\alpha>3/2$ the typical difference of the frequency sums exceeds the maximum interaction sum, and so synchronization of separated blocks would not be expected for finite $K$. Comparing the scaling of the interaction and frequency sums with $M$ suggests that an interaction strength scaling with system size $N$ as $N^{\alpha-3/2}$ is required for macroscopic synchronization for $\alpha > 3/2$. This reduces to the result $K\sim N^{1/2}$ for $\alpha=2$ where the block sums reduce to the nearest-neighbor results, consistent with the rigorous result of Strogatz and Mirollo \cite{strogatz} for the nearest neighbor model.

\section{Numerical Simulations}
\label{num}
The numerical evolution of Eq.\ (\ref{longmod}), has been carried out for five 
different system sizes, namely $N=256$, $1024$, $4096$, $8192$ and $16384$. Periodic boundary conditions were imposed.

In order to integrate the equations of motion, we have used the fourth order 
Runge-Kutta method with a time step typically $\Delta t=0.05$. For the long range model with a system size $N$, there are $^{N}C_2 $ 
interaction terms. Therefore, evaluation of the interaction kernel using the 
direct method in real space would be an ${\cal{O}}(N^2)$ operation at each time step. Instead we express the interaction as a convolution
\begin{equation}
\sum_{i\neq j}K_{ij}\sin(\theta_{i}-\theta_{j})=\operatorname{Im}[e^{-i\theta_{j}}
\sum_{i\neq j}K(|i-j|)e^{i\theta_{i}}]
\end{equation}
which can be efficiently evaluated in ${\cal{O}}(N\log N)$ operations using fast Fourier transforms.

The initial phases are randomly distributed between $-\pi$ 
and $\pi$. The intrinsic frequencies are Gaussian random numbers with zero mean
 and unit variance.
The first $t_0=200$ time units are discarded as transients while integrating the 
equations. The data is then accumulated for $t_f-t_0$ additional 
time units. We use $t_f-t_0=600$ for most of the results, but increase this to 2400 to analyze the synchronized clusters in Section \ref{Sec_clusters}. The phase order parameter 
has been time averaged over the entire range $t_f-t_0$, while the Edwards-Anderson order 
parameter has been calculated as $\Psi_{EA}=|\sum_{j=1}^{N}e^{i(\theta_{j}(t_f)-\theta_{j}(t_{0}))}|/N$.  The data for the order
parameters and the correlation functions have also been averaged over $30$ different
initial configurations, by repeating with different seeds for the random number generator.
We present the results for a coupling strength $K=1$.

\subsection{Order Parameters}

\begin{figure}[tbh]
\begin{center}
\includegraphics[angle=-90,width=0.9\columnwidth]{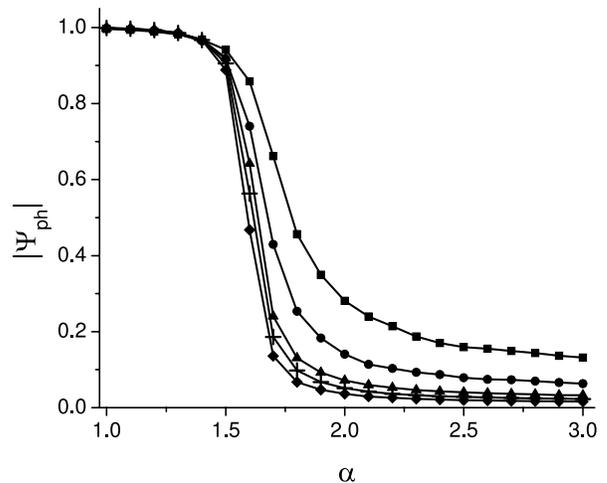}
\end{center}
\caption{The phase order parameter $|\Psi_{ph}|$ as a function of the power law $\alpha$ of the interaction for five different system sizes. The strength of the coupling is $K=1$. The system size increases from the upper curve to the lower: squares - 256; circles - 1024; triangles - 4096; pluses - 8192; diamonds - 16384.}
\label{opph}
\end{figure}

\begin{figure}[tbh]
\begin{center}
\includegraphics[width=0.9
\columnwidth]{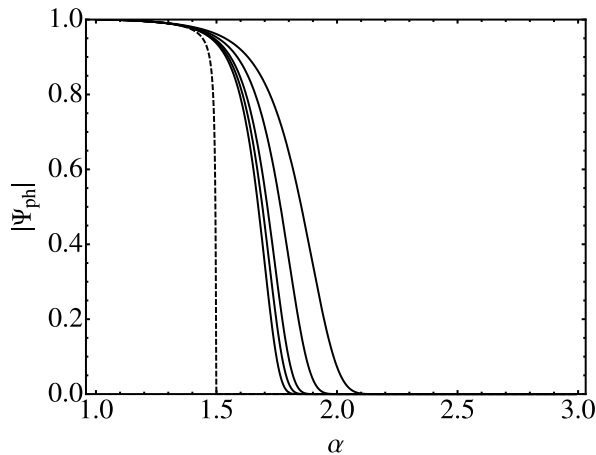}
\end{center}
\caption{Estimate of the phase order parameter $|\Psi_{ph}|$ (solid curves) as a function of the power law $\alpha$ of the interaction based on Eqs.\ (\ref{Delta theta N/2}) and (\ref{Psi N estimate}) for the same five system sizes used in Fig.\ \ref{opph}, increasing from the upper curve to the lower. The strength of the coupling is $K=1$ and a value of $\beta=0.5$ was used. The spin wave prediction for an infinite system size is shown for reference (dashed curve).}
\label{opes}
\end{figure}

We show the magnitude of the phase order parameter Eq.\ (\ref{define OP}) as a function of the power law $\alpha$ of the interaction, for coupling strength $K=1$ and for the five different system sizes in Figs.\ \ref{opph}.  The order parameter decreases rapidly for a value of $\alpha$ that decreases with increasing system size, with the half-height $|\Psi_{ph}|=0.5$ value occurring at about $\alpha=1.6$ for the largest system size used.
As we have seen in Section \ref{Section Order Parameter} we expect strong finite size effects for $\alpha$ near 3/2, so that it is hard to extrapolate the numerical data to the infinite size limit. For comparison we show in Fig.\ \ref{opes} the finite size estimate of the order parameter from Eqs.\ (\ref{Psi N estimate}) and (\ref{Delta theta N/2}). We do not expect a quantitative agreement for any $\alpha$ due to the crude approximations made, and the behavior for small $|\Psi_{ph}|$ is not correct as explained in Section \ref{Section Order Parameter}, but the overall trends are quite similar.

\begin{figure}[tbh]
\begin{center}
\includegraphics[angle=-90,width=0.8\columnwidth]{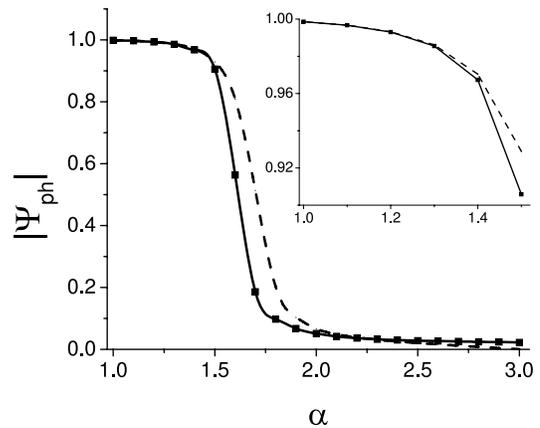}
\caption{Comparison between the spin-wave sum (dashed line) and the simulation results (squares joined by solid line) for the phase order parameter $|\Psi_{ph}|$ as a function of the power law $\alpha$ of the interactions in a system of size 8192 and for coupling strength $K=1$. The inset shows the same comparison over a smaller range}
\label{opcomp}
\end{center}
\end{figure}

A comparison between the full spin-wave predictions, performing the discrete sums without approximations, and the results from the simulations of the time evolution, both for system size 8192, is shown in Fig.~{\ref{opcomp}}. 
The plot shows quantitative agreement between the simulations and the spin wave sums 
for $\alpha\leq 1.4$. As $\alpha$ approaches $1.5$ and the order parameter decreases, there is
increasing disagreement, as would be expected since the theory is an expansion assuming well aligned phases. 

Although the strong finite size effects preclude reliably extrapolating to infinite system sizes, the numerical simulation results appear to be consistent with the prediction of the spin wave theory, that phase ordering and entrainment is possible for $\alpha<3/2$, but that phase order is not possible for $N\to\infty$ for $\alpha>3/2$.

\begin{figure}[tbh]
\begin{center}
\includegraphics[angle=-90,width=0.9\columnwidth]{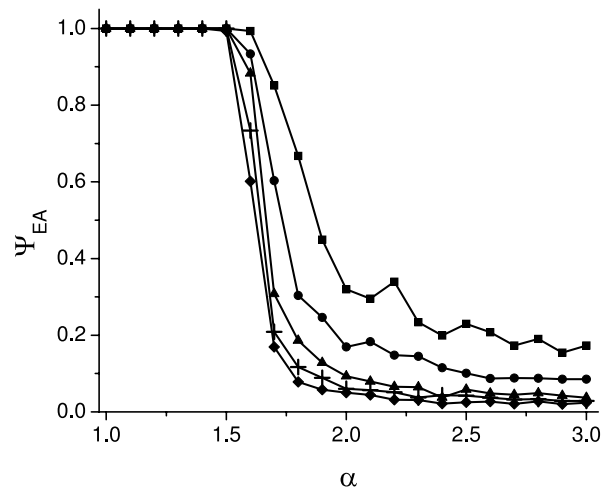}
\end{center}
\caption{The Edwards-Anderson order parameter $\Psi_{EA}$ as a function of the power law $\alpha$ of the interaction for five different system sizes as in Fig.\ \ref{opph} and coupling strength $K=1$.}
\label{opea}
\end{figure}

The Edwards-Anderson order parameter evaluated from the simulations for the same parameter values is shown in Fig.\ \ref{opea}. For all system sizes used and coupling strength $K=1$, $\Psi_{EA}$ remains unity up to $\alpha=3/2$, showing perfect entrainment, suggesting a single block of all the oscillators evolving at the mean frequency of the distribution for the system sizes used. For $\alpha >3/2$ the order parameter decreases, showing that the system has broken up into more than one synchronized cluster. Qualitatively, the fall off above $\alpha=3/2$ is similar to the fall off of the phase order parameter. Thus we do not find stronger correlations in the oscillator frequencies than in the phases. This does not appear to be consistent with the predictions from the spin-wave theory of a state with entrained oscillators but without long range phase order for the range $3/2 < \alpha <5/2$. On the other hand the result is consistent with the block sum arguments which show that for $\alpha < 3/2$ blocks of oscillators may synchronize across regions of unsynchronized oscillators, whereas this does not typically happen for $\alpha > 3/2$. 

\subsection{Correlation Function}

\begin{figure}[tbh]
\begin{center}
$\begin{array}{c@{\hspace{.1in}}c@{\hspace{.1in}}c}
\includegraphics[angle=-90,width=1.6in]{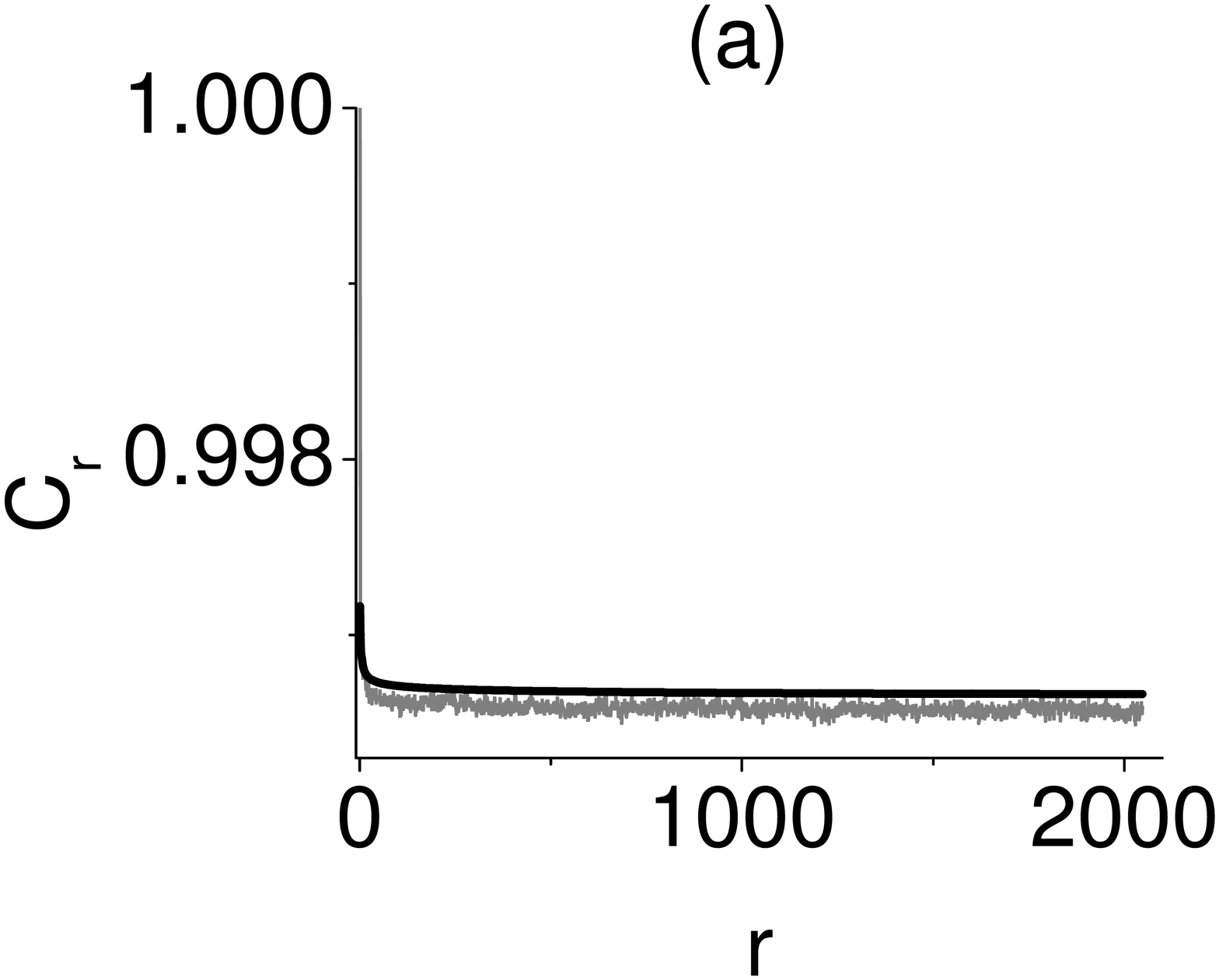}  &  
\includegraphics[angle=-90,width=1.6in]{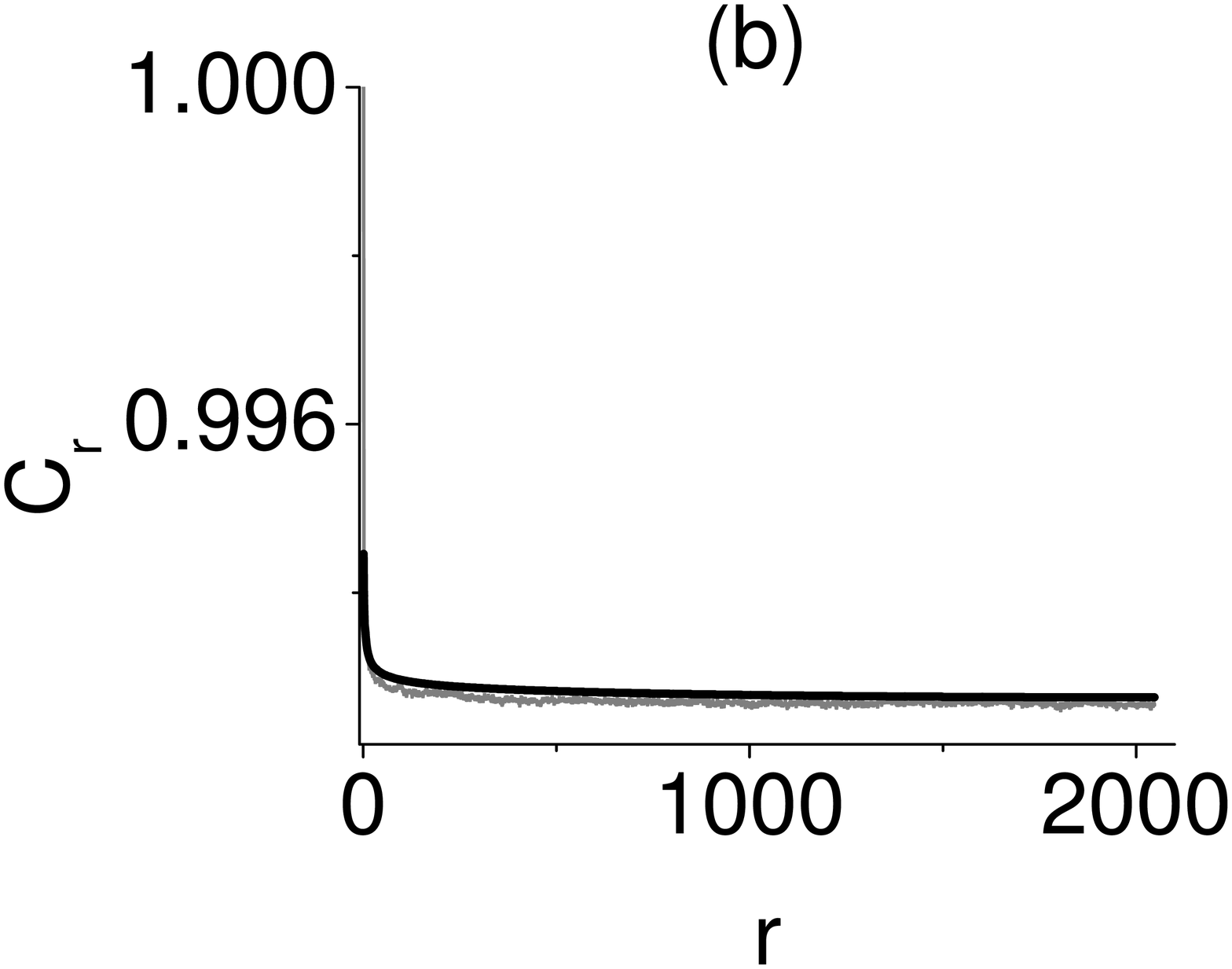} \\
\includegraphics[angle=-90,width=1.6in]{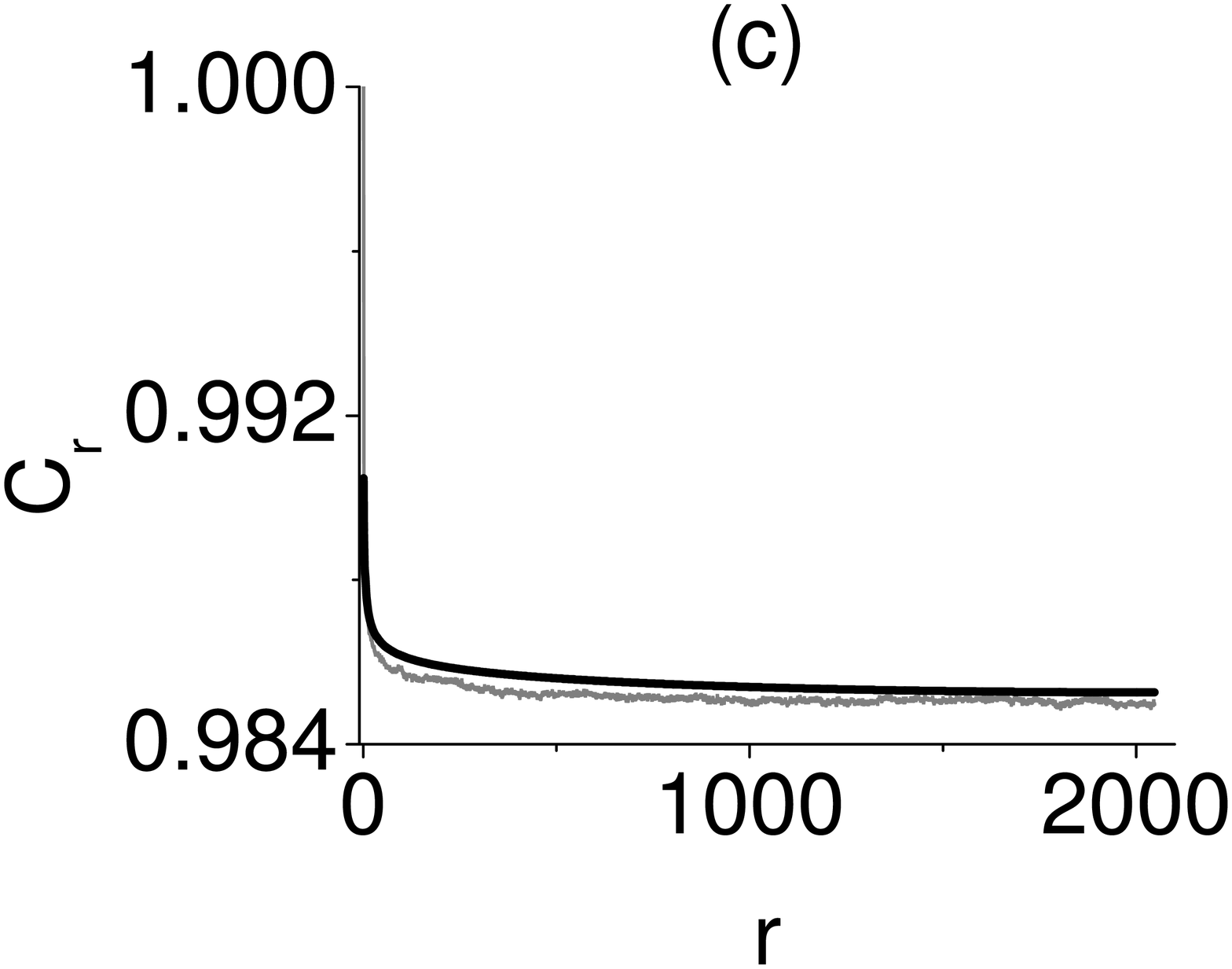} & 
\includegraphics[angle=-90,width=1.6in]{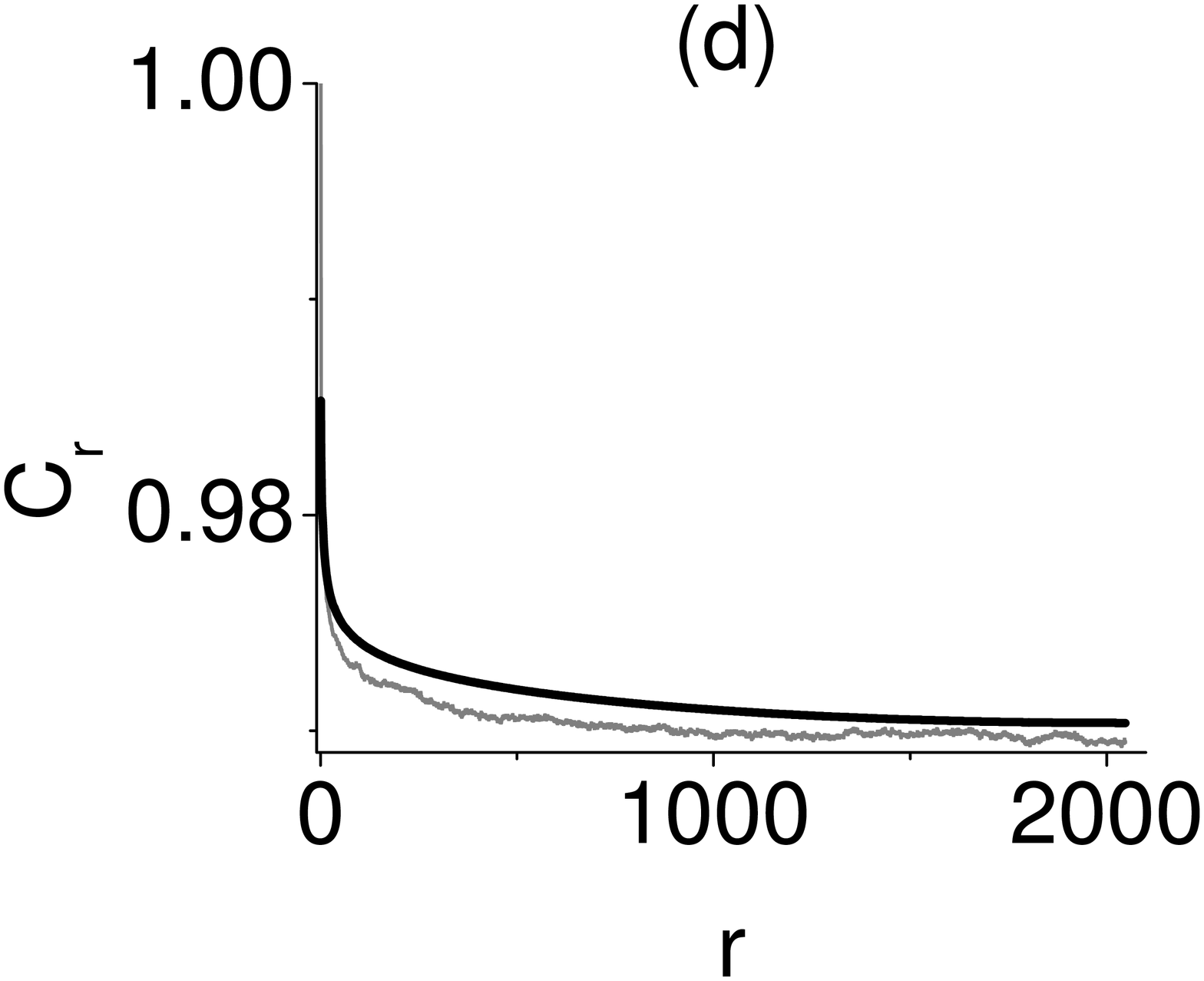} \\   
\includegraphics[angle=-90,width=1.6in]{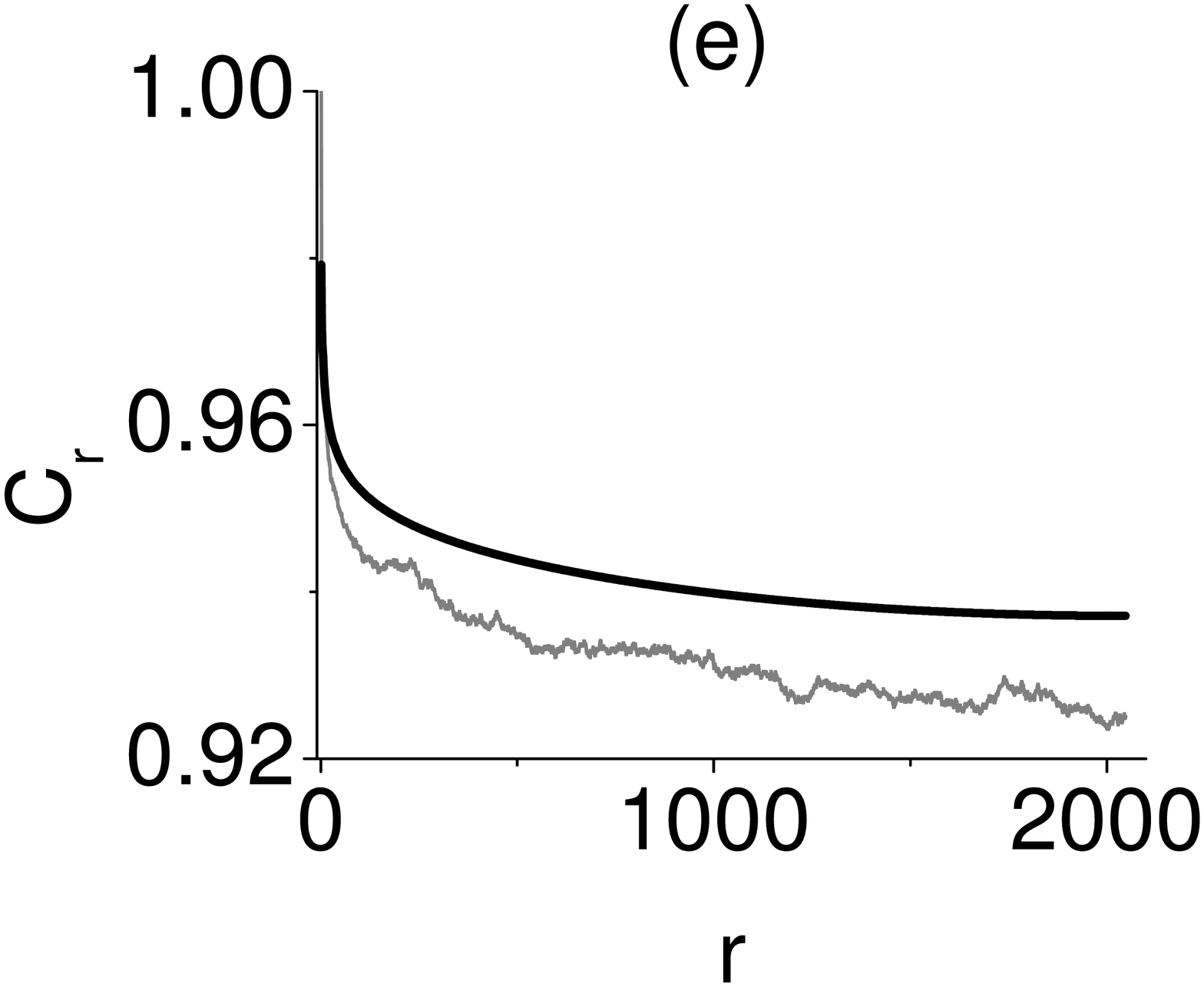} &
\includegraphics[angle=-90,width=1.6in]{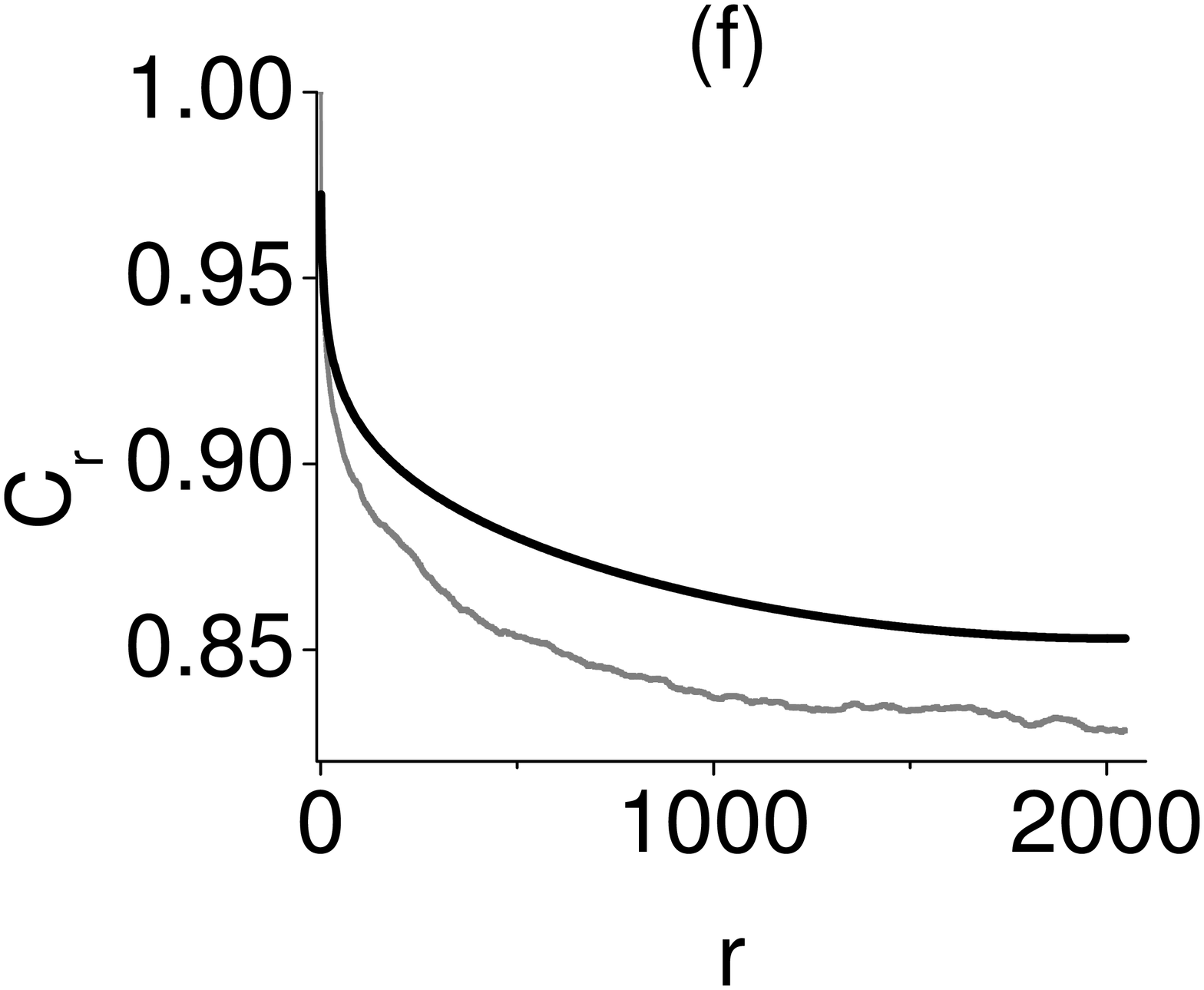} \\
\end{array}$
\end{center}
\caption{Comparison between the spin-wave sum and the simulation results for the correlation function $C_{r}$ as a function of separation $r$ for six values of the interaction power law exponent $\alpha$ and a system size of $4096$ and coupling strength $K=1$: dark lines - spin wave prediction; light points - simulation results.}
\label{fig}
\end{figure}

As a further test of the results of the spin wave approach we show in Fig.\ \ref{fig} a comparison between the phase correlation function Eq.\ (\ref{correlation function}) evaluated from the spin wave discrete sums and the full numerical simulations for six values of $\alpha$ in the range $1\leq\alpha\leq 3/2$ for a system size 4096.
The plots show quantitative agreement for $\alpha\leq1.3$. As $\alpha$ approaches $1.5$, the difference grows, which is consistent with what we saw  earlier in Fig.\ {\ref{opea}}.
 
\begin{figure}[tbh]
\begin{center}
\includegraphics[angle=-90,width=0.8\columnwidth]{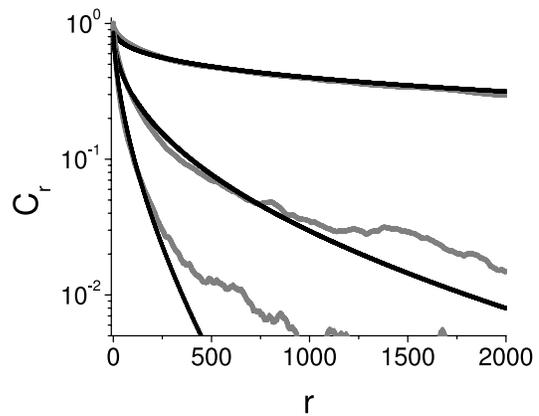}
\end{center}
\caption{Phase correlation function as a function of separation for system size 8192: light points - simulations; dark lines - stretched exponential fits, Eq.\ (\ref{Stretched exponential}). The values of the interaction power law $\alpha$ are, top to bottom, $\alpha=1.6$, $\alpha=1.7$, and $\alpha=1.8$.}
\label{compcorse}
\end{figure}
 
For $3/2 < \alpha < 2$ the spin wave analysis predicts a phase correlation function with a stretched exponential decay, see Eq.\ (\ref{C cases}).  Based on these predictions, we fit the correlation function obtained from the simulations in this range of $\alpha$ to a function
\begin{equation}
C(r)=e^{-br^c}\label{Stretched exponential}
\end{equation}
for $\alpha=1.6$, $\alpha=1.7$, and $\alpha=1.8$ and system size 8192, see Fig.\ \ref{compcorse}. An exponential correlation would be a straight line on these log-linear plots, and clearly does not fit the data. The fit parameters for the stretched exponential and the 
predictions of the spin wave theory are shown in Table \ref{tablefit}. 
\begin{table}
\begin{tabular}{|l|l|l|l|l|}
\hline
&b(Numerical)&b(SW)&c(Numerical)&c(SW)\\
\hline
$\alpha=1.6$ & $0.094\pm 0.001$ & $0.093$& $0.330\pm  0.001$ & $0.2$\\
$\alpha=1.7$ & $0.151\pm 0.001$ & $0.064$ &$0.456\pm 0.001$ & $0.4$\\
$\alpha=1.8$ & $0.151\pm  0.001$&$0.055$ & $0.583\pm 0.002$ & $0.6$\\
\hline
\end{tabular}
\caption{\label{tablefit} Parameters of the stretched exponential correlation function Eq.\ (\ref{Stretched exponential}) from fits to numerical results for system size 8192 and three values of $\alpha>1.5$ compared with predictions from spin-wave theory for the infinite system.}
\end{table}
The agreement between the exponents is reasonably good for $\alpha=1.7$ and $\alpha=1.8$. The deviation from the predictions for $\alpha=1.6$ can be ascribed to finite size effects, since the correlations tend to a finite value for large $r$ corresponding to separations of $N/2$ for this value of $\alpha$, and so should not be compared with the theoretical predictions for an infinite size system. In the next section we discuss the size of the synchronized clusters. The range over which the stretched exponential fit is good in Fig.\ \ref{compcorse} is comparable to the maximum cluster size.

\subsection{Clusters}
\label{Sec_clusters}
In this section we present results from our simulations for the size of synchronized clusters as a function of the power law of the interaction. For long range interactions both contiguous blocks, and disjoint blocks entrained through the long range interaction across unentrained oscillators, are of interest.

We identify an entrained cluster from the simulations as a set of oscillators that are phase locked: over the time of the simulation no oscillator phase undergoes slips (changes of about $\pm2\pi$) with respect to the mean phase. In a simulation over a time $T$ this is equivalent to the frequency being within $2\pi/T$ of the mean frequency of the cluster. The expected frequency difference between large but distinct clusters of size about $L$ is of order $L^{-1/2}$. Thus the simulation time should exceed $2\pi L^{1/2}$.
We compute the phase-winding number, $n_w$, for every oscillator along the 
chain. The phase-winding number is calculated as
\begin{equation}
n_w=2\left[\frac{\lim_{t-t_0\to\infty}(\theta_i(t)-\theta_i(t_0))}{4\pi}\right],
\label{pwno}
\end{equation}
where $[x]$ denotes the nearest integer to $x$.

\begin{figure}[tbh]
\begin{center}
\includegraphics[width=0.8\columnwidth]{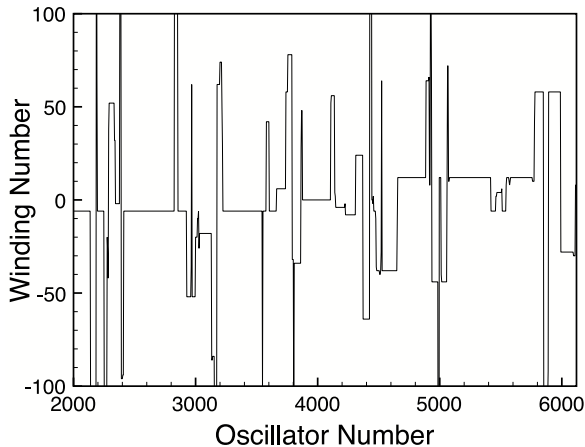}
\end{center}
\caption{Winding numbers over a portion of the system for $\alpha=1.7$ and $K=1$ and a system size of $8192$.}
\label{freqplat}
\end{figure}

An example of the raw data of winding numbers, computed over a run time $T=t-t_{0}$ of 2400, is shown in Fig.\ \ref{freqplat} for a system size $N=8192$, coupling strength $K=1$ and interaction power law $\alpha=1.7$, Only a portion of the full system is shown. Note in particular blocks with the same winding numbers entrained across oscillators with different winding numbers, a novel consequence of the long range interaction.

\begin{figure}[tbh]
\begin{center}
$\begin{array}{c@{\vspace{.1in}}c@{\vspace{.1in}}c}
\includegraphics[angle=-90,width=1.7in]{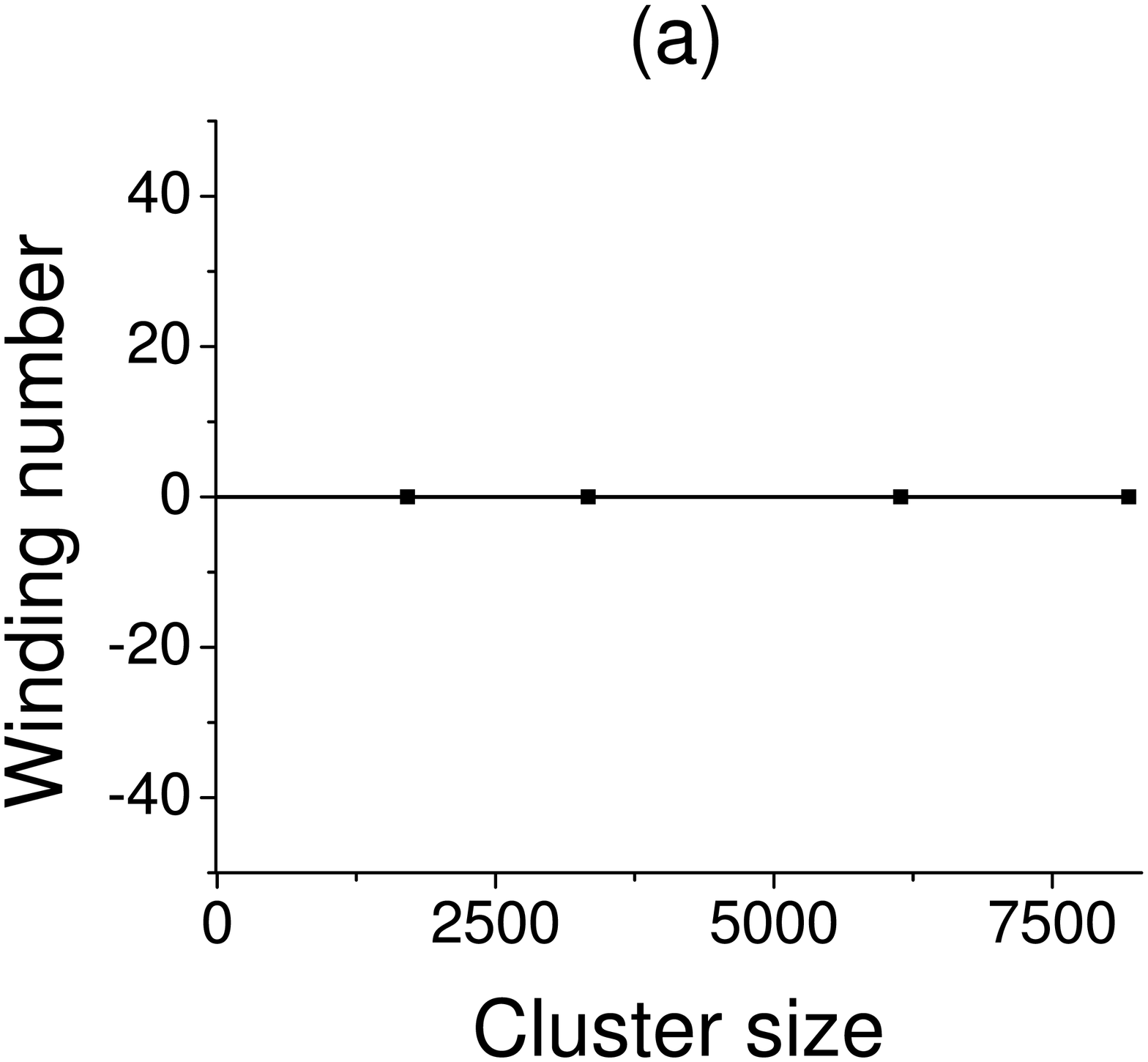}  &  
\includegraphics[angle=-90,width=1.7in]{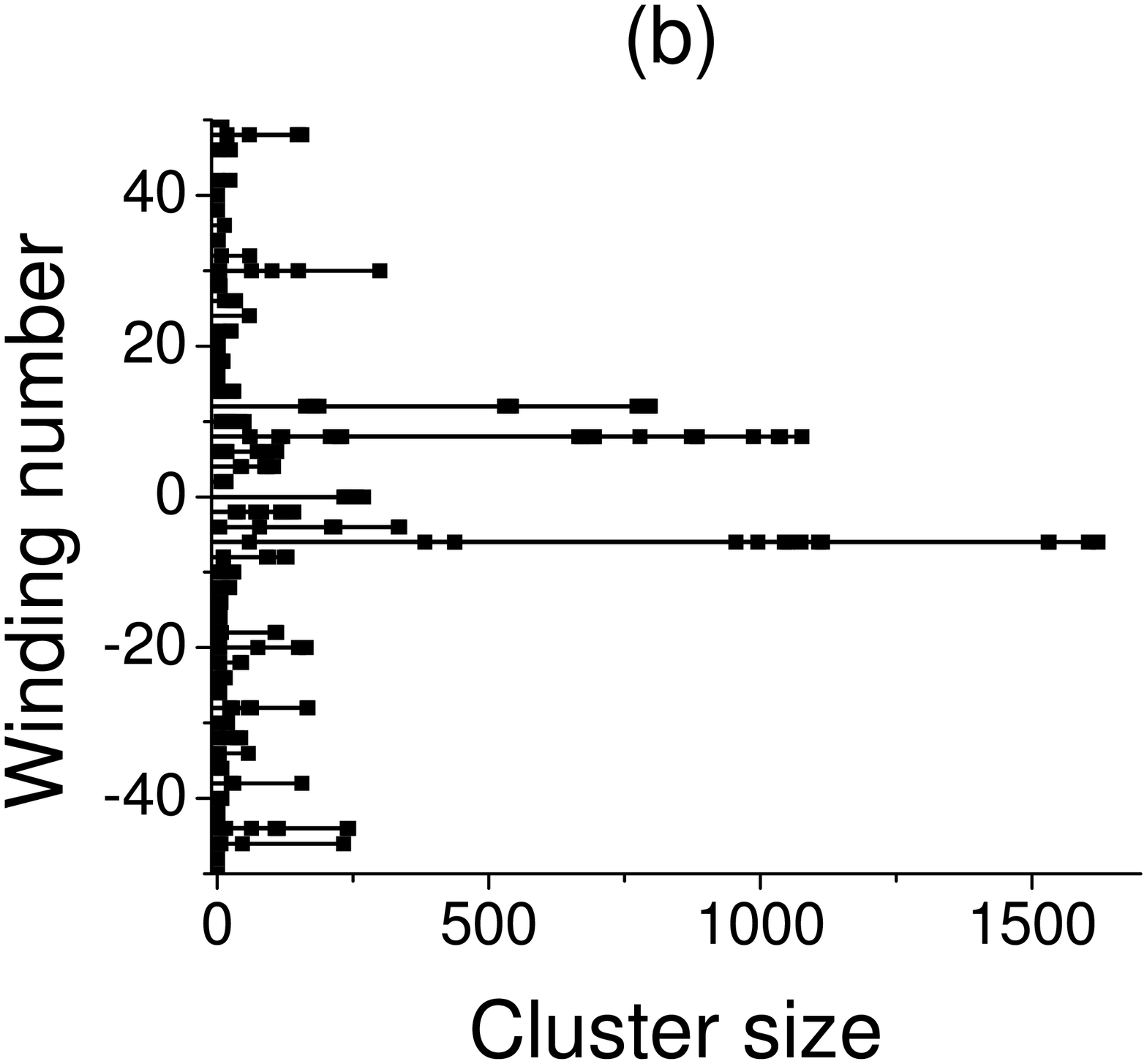} \\
\includegraphics[angle=-90,width=1.7in]{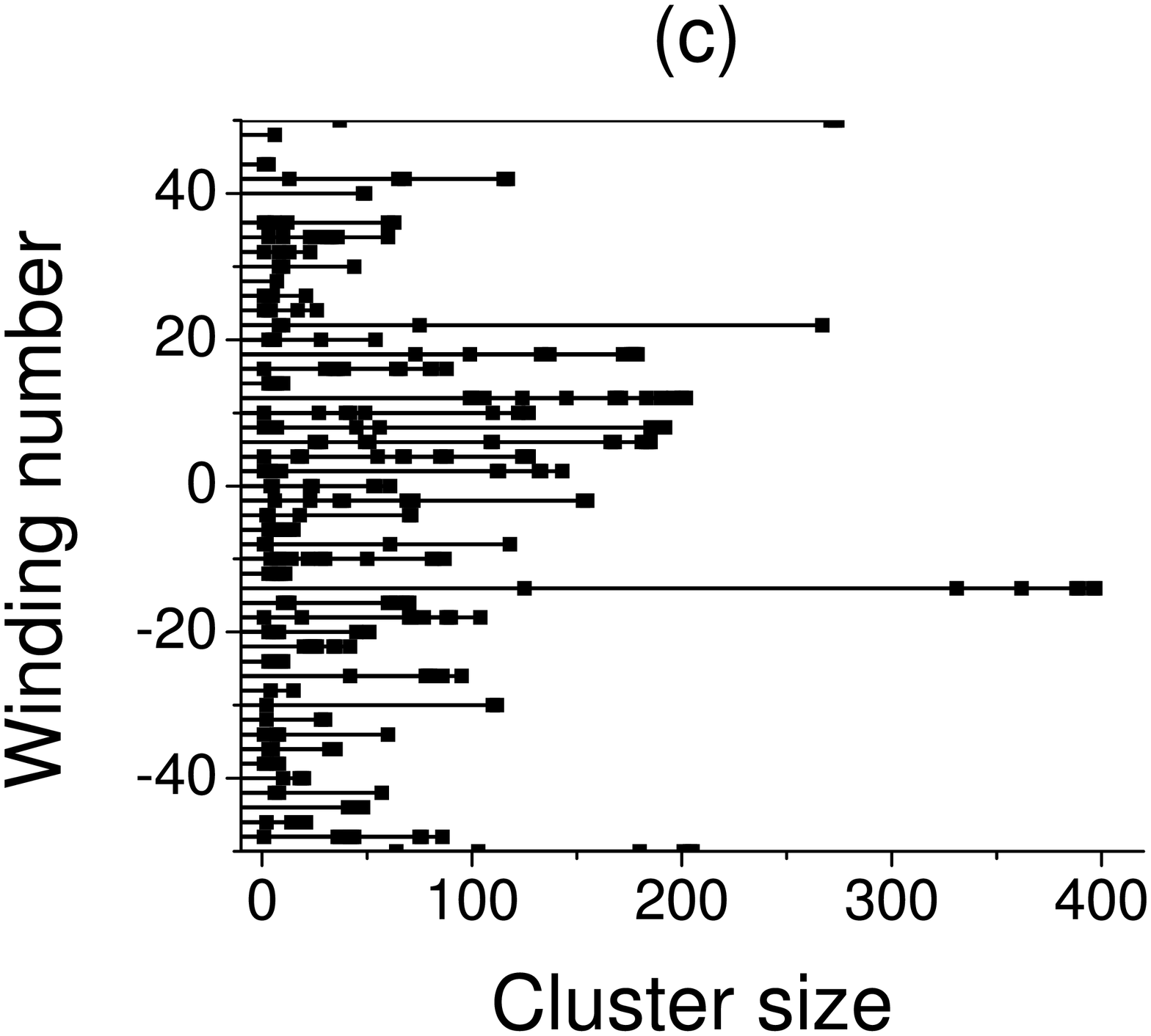}  & 
\includegraphics[angle=-90,width=1.7in]{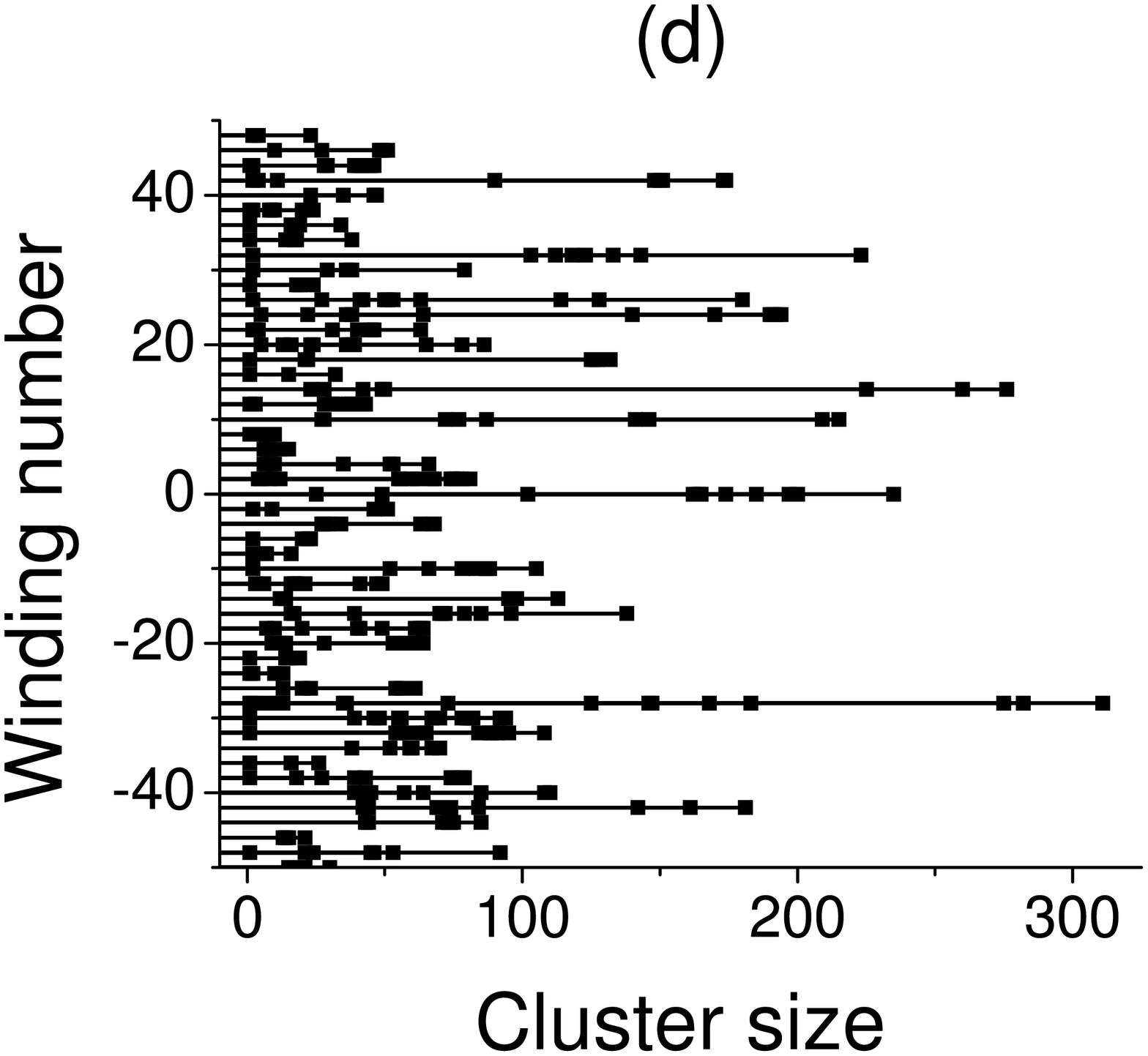} \\  
\end{array}$
\end{center}
\caption{Phase winding numbers $n_{w}$ as a function of cluster size for
 a system of size $8196$ and interaction strength $K=1$. The lengths of the bars define the total number of oscillators  with the same winding number, and the lengths between the points on the  bars give the sizes of the individual contiguous blocks making up the whole. The four plots are for different  interaction power laws: (a) $\alpha=1.6$; (b) $\alpha=1.7$; (c) $\alpha=1.8$; (d) $\alpha=1.9$.
 Only a restricted range of winding number is shown.}
\label{histo}
\end{figure}
Histograms of the phase winding number for the same system are shown in Fig.\ {\ref{histo}} for four values of $\alpha$. We use the total number of oscillators with the same winding number, shown by the bar length, to define the overall cluster size, and this is divided up into the individual contiguous blocks (containing no oscillators with different winding numbers) given by the lengths between the points on the bars. For clarity, only a restricted range of winding number is shown: there are additional small clusters with more distant winding numbers outside the range plotted. For $\alpha=1.6$ almost all the oscillators have the same winding number $n_{w}=0$, so that the cluster of entrained oscillators spans the whole system, with only a few oscillator of different winding numbers breaking the global cluster into four smaller contiguous blocks. As $\alpha$ increases, more clusters of smaller size develop. In each case a number of contiguous blocks join to form a large cluster with entrainment across unentrained oscillators.
\begin{figure}[tbh]
\begin{center}
\includegraphics[angle=-90,width=3.0in]{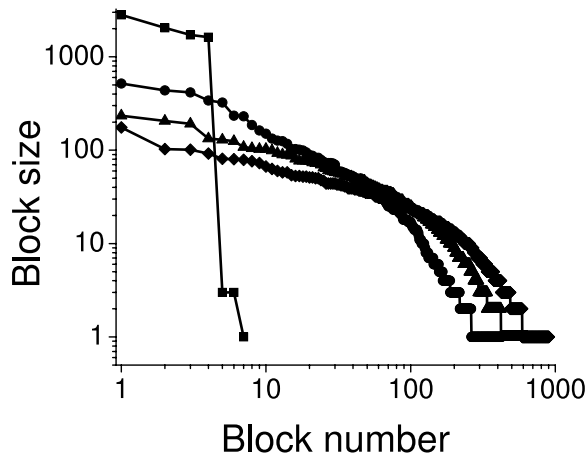}
\caption{Ordered plot of contiguous block sizes for four values of the power law $\alpha$: squares - $\alpha=1.6$; circles - $\alpha=1.7$; triangles - $\alpha=1.8$; diamonds - $\alpha=1.9$. The ordinate is the size of a contiguous entrained block, and the abscissa the index number in the list of blocks ordered by size. Other parameters are as in Fig.\ \ref{histo}.}
\label{cluster distribution}
\end{center}
\end{figure}

The full distribution of contiguous block sizes is shown in Fig.\ \ref{cluster distribution}. This is a plot of an ordered list of the sizes of contiguous blocks. A log-linear plot of the same data shows a good fit to an exponential fall off for small block sizes (the large block number end of the plot).


\section{Conclusions}

We have studied the synchronization of oscillators described by a phase only model with interactions falling off with separation $r$ as a power law $r^{-\alpha}$ using an expansion about the aligned phase state (spin-wave method), arguments summing the equations of motion of blocks of oscillators (block-sum method) and numerical simulations on systems of up to 16384 oscillators. We have focussed on the range $\alpha\geq 1$, since previous work \cite{marodi} has looked at $\alpha<1$ in some detail.

For $1 \leq\alpha<3/2$ we find results consistent with macroscopic entrainment and long range phase order for large enough coupling strengths. The spin-wave type analysis, based on the assumption of a time independent solution (fully entrained state) and an expansion of the interaction term linearly in $\theta_{i}-\theta_{j}$, predicts a state with long-range phase order and a nonzero phase order parameter. For large enough coupling strength $K$ the mean square phase deviation $\langle (\theta_{i}-\theta_{j})^{2}\rangle$ is small, and the average phase correlations weighted by the power law interaction are close to unity, so that the linear expansion of the nonlinear interaction function is a good approximation on average. The block sum argument shows that despite the long range interaction, the coupling of a finite block to the rest of the chain remains bounded above by some finite value, and there is a nonzero probability of finding a finite block of oscillators with frequencies sufficiently far from the mean that they are not synchronized to the rest of the chain for finite coupling.
This argument shows that for $N\to\infty$, there are no macroscopic ($O(N)$) contiguous blocks of synchronized oscillators for any finite $K$, so that the assumption of a time independent solution as made in the spin-wave approach is not correct \footnote{The reconciliation of this result with the spin-wave calculations presumably lies in calculating the full distribution of $\theta_{i}-\theta_{j}$ and finding tails of the distribution leading to some probability of values of order unity even for large $K$.}. However, for $\alpha<3/2$, the interaction is sufficiently long range that large blocks of oscillators are likely to synchronize {\it across} the unsynchronized oscillators (see Fig.\ \ref{freqplat} for examples from the simulations), leading to the entrainment of a finite fraction of the oscillators, long range phase correlations, and a nonzero order parameter for sufficiently large $K$, even for $N\to\infty$. These results follow the predictions of the spin-wave theory, although the finite blocks of unsynchronized oscillators will reduce the order parameter and phase correlations below the value predicted by spin wave theory, as seen in the comparison of the simulations with the spin-wave predictions.

For $3/2<\alpha<5/2$ the spin-wave approach predicts a fully entrained state, but with no long range phase order, although for $\alpha<2$ the phase correlations are predicted to be of stretched-exponential form. However the block sum method shows that finite unsynchronized blocks again exist, and now large blocks of oscillators will typically not synchronize across the unsynchronized oscillators. Thus for finite coupling strength $K$ and $\alpha>3/2$ we expect {\it no} macroscopic entrainment (no finite fraction of oscillators at the same frequency for $N\to\infty$). The results of the numerical simulations show the Edwards-Anderson order parameter which measures frequency entrainment, decreasing as $\alpha$ increases above 3/2 in a way that is broadly similar to the phase order parameter, consistent with the picture that the unsynchronized blocks disrupt both the phase and frequency correlations. The simulations show results consistent with the spin-wave predictions of a stretched exponential decay of correlations up to a distance comparable with the largest cluster size.

Rogers and Wille concluded in their paper that the  critical interaction exponent $\alpha_{c}$
such that the oscillators do not synchronize for $\alpha>\alpha_{c}$ even for very large
coupling strengths is $\alpha_{c}\simeq 2$. Our results suggest a lower critical value of $\alpha_{c}=3/2$ and our numerics on larger systems than used in ref.\ \cite{rogers} approach this value. The diagnostic used by Rogers and Wille was the average plateau size  as a fraction of the system size. In their simulations they found this quantity to switch quite rapidly as a function of increasing $\alpha$ for reasonably large $K$ from unity to close to zero. The plateaus were defined as contiguous blocks of oscillators with the same frequency, and so oscillators synchronized across
unsynchronized blocks were not counted as in the same plateau. This means that their diagnostic does not detect long range synchronization occurring through this mechanism. However, we believe the main reason that their value of $\alpha_{c}$ is greater than the value 3/2 that we propose is due to the strong finite size effects for $\alpha$ near 3/2, so that for the range of sizes they used, too large a value of $\alpha$ is needed for desynchronization to appear, and their extrapolation scheme to large $N$ was not adequate.

Marodi et al. have also looked at this problem numerically for sizes up to $1000$
 in one dimension (as well as two dimensions). The main focus of their work was $\alpha<1$, where complete entrainment occurs for $N\to\infty$ for the scaling of the coupling constant they use. They do not attempt to identify a critical value of $\alpha>1$ above which partial synchronization is no longer possible. Their numerics on system sizes up to 1000 and for $K=1$ show the phase order parameter decreasing for $\alpha$ close to 3/2 --- in fact closer than we find for these system sizes (compare their Fig.\ 2 with our Fig.\ \ref{opph}) perhaps because of the open boundary conditions they use. They also remark that the order parameter approaches a steady value for $\alpha\gtrsim 2.5$, but we believe this value tends to zero for large $N$, which is consistent with the trends in their numerics.

\acknowledgements

We thank Gil Refael, Tony Lee and Hsin-hua Lai for interesting 
discussions. DC thanks the IIT Kanpur-Caltech MoU and the SURF program at Caltech for 
financial support.


\begin{thebibliography}{99}
\bibitem{Pikovsky} A. Pikovsky, M. Rosenblum and J. Kurths, {\it Synchronization: a Universal Concept in Nonlinear Sciences} (Cambridge University Press, Cambridge, England, 2001).
\bibitem{bargatin} I. Bargatin, Ph.D. Thesis Caltech, 2008.
\bibitem{winfree} A. T. Winfree,  J. Theor. Biol. {\bf16}, 15 (1967).
\bibitem{kurabook} Y. Kuramoto, {\it Chemical Oscillations, Waves and Turbulence}, (Dover, New York, 2003).
\bibitem{acebron} J. A. Acebr$\acute{o}$n et al., Rev. Mod. Phys. {\bf 77}, 137 (2005).
\bibitem{daido} H. Daido, Phys. Rev. Lett. {\bf 61}, 231 (1988).
\bibitem{strogatz} S. H. Strogatz and R.E. Mirollo, J. Phys. A {\bf 21}, L699 (1988); S. H. Strogatz and R.E. Mirollo, Physica D {\bf 31}, 143 (1988).
\bibitem{sakaguchi_1} H. Sakaguchi, S. Shinomoto, and Y. Kuramoto, Prog. Theor. Phys. {\bf 77}, 1005 (1987)
\bibitem{hong} H. Hong, H. Park, and M. Y. Choi, Phys. Rev. E {\bf 72}, 036217 (2005).
\bibitem{rogers} J. L. Rogers and L. T. Wille, Phys. Rev. E {\bf 54}, R2193 (1996).
\bibitem{marodi} M. Mar$\acute{o}$di, F. d'Ovidio and T. Vicsek, Phys. Rev. E {\bf 66}, 011109 (2002).
\bibitem{pnas} P. K{\"o}nig, A. K. Engel and W. Singer, Proc. Natl. Acad. Sci.\ {\bf 92}, 290 (1995).
\bibitem{radicchi} F.\ Radicchi and H.\ Meyer-Ortmanns, Phys. Rev. E {\bf 74}, 026203 (2006).
\bibitem{kuraprog} H. Sakaguchi, S. Shinomoto and Y. Kuramoto, Prog. of Theor. Phys. {\bf 77}, 1005 (1987).
\bibitem{tasaki} T. Koma and H. Tasaki, Phys.\ Rev.\ Lett.\ {\bf 74}, 3916 (1995).
\end{thebibliography}
\end{document}